\documentclass[preprint,pre,aps,superscriptaddress,a4paper]{revtex4}
\bibstyle{prsty}
\usepackage[utf8x]{inputenc}
\usepackage{graphicx, 
}
\usepackage{epstopdf}
\usepackage{amsmath}
\usepackage{amsfonts}
\usepackage{color}
\begin{document}
	\title{Correlation functions in mixtures with energetically favoured nearest-neighbours of different kind: a size-asymmetric case}
	\author{ O. Patsahan}
	\affiliation{Institute for Condensed Matter Physics of the National
		Academy of Sciences of Ukraine, 1 Svientsitskii St., 79011 Lviv,
		Ukraine}
	\author{ A.  Meyra}
	\affiliation{IFLYSIB (UNLP, CONICET), 59 No. 789, B1900BTE La Plata, Argentina;  Depto de Ingenieria Mec\'{a}nica, UTN-FRLP, Av. 60 esq. 124, 1900 La Plata, Argentina}
	\author{ A. Ciach}
	\affiliation{Institute of Physical Chemistry,
		Polish Academy of Sciences, Kasprzaka 44/52, 01-224 Warszawa, Poland}
	\date{\today} 
	\begin{abstract}
	Binary mixtures of hard-spheres with different diameters and square-well attraction between different particles are studied by theory and Monte Carlo simulations.  
	In our mesoscopic theory, local fluctuations of the volume fraction of the two components are taken into account. Semi-quantitative agreement between  the simulation and 
	theoretical results is obtained, except from very small distances. The correlation functions exhibit exponentially damped oscillations, with the period  determined by the
	interaction potential, and both the amplitude and the correlation length increasing significantly with increasing diameter ratio. Increasing size asymmetry leads also to decreasing 
	fluctuations of the number of  the smaller particles in the attractive shell of the bigger ones.  For small size asymmetry, the strongest correlations occur for comparable volume 
	fraction of the two components. When the size ratio increases, the maximum of the structure factor moves to a larger volume fraction of the bigger particles, and for the size ratio as
	large as 4, the maximum goes beyond the accessible range of volume fractions. Our results show that when the neighbourhood of different particles is energetically favoured, 
	the particles are much more uniformly distributed than in the random distribution even at relatively high temperature, especially for large size asymmetry. 
	
	\end{abstract}

\maketitle
\section{Introduction}

Local structure in simple liquids is determined by packing of hard spheres representing particle cores.  Even in the case of 
spherical shapes, however, the distribution of 
the particles  or molecules can be significantly different 
when the interactions between them exhibit some competing tendencies.
In particular,
in the case of electrostatic interactions, repulsion between like charges competes with attraction of opposite charges. 
At sufficiently low temperature,
this competition leads to various ionic crystals in simple salts, as well as
 to a rich variety of ordered structures in mixtures of oppositely charged colloid particles with different size 
 ratios~\cite{royall:06:0,blaaderen:05:0,hynninen:06:0}.
 In the fluid phase, the order can still be present locally, as observed in simulations of ionic liquids (IL)
 and IL mixtures (ILM)~\cite{Shimizu2015}, and predicted 
 theoretically~\cite{stillinger:68:0,leote:94:0,ciach:05:0,ciach:05:2,ciach:06:2,patsahan:12:0,patsahan:07:0}.
 The local order can be directly observed near a planar boundary, since the concentration
 or density profiles are similar to the corresponding correlation functions. 
 Indeed, charge density oscillations in direction perpendicular to 
 the planar boundary have been found in experiments, simulations and theory~\cite{smith:16:0,fedorov:08:0,fedorov:14:0,otero:18:0,ciach:18:1}.
 In directions parallel to the confining plane,
 either hexagonal or stripe pattern formed by coions is formed  in  the near-surface layer, 
 and the order extends beyond the first two layers of ions in ILM~\cite{montes:17:0,otero:18:0}. 
 Local order in mixtures of colloid particles can be induced by non-coulombic interactions as well, for example by H-bonds formation between 
 different particles~\cite{bradley:11:0}.
 
 Determination of the local structure in the case of Coulombic interactions by traditional liquid theories
 is very difficult~\cite{leote:94:0,attard:93:0}, especially in the case of large size asymmetry.
 It is easier to determine the structure of the fluid within
 the density functional theory (DFT)~\cite{evans:79:0} that can very well predict the structure induced by the packing of hard spheres. 
 Unfortunately, in the standard DFT 
 the interactions are taken into account in an expression of the mean-field (MF) type.
 This expression is  inaccurate in a disordered phase with the local structure induced by the competing interactions, 
 and the internal energy 
 in this phase is overestimated~\cite{ciach:18:0}. As a result, instead of a disordered phase with strong local order, 
 an ordered phase with 
 long-range charge oscillations with a small amplitude is predicted for a significant part of 
 the phase diagram~\cite{ciach:03:1,ciach:05:0,ciach:07:0}. 
   A continuous transition to this ordered phase occurs at a $\lambda$-surface, where
 the correlation functions in Fourier representation diverge for a particular wavenumber~\cite{ciach:05:0,ciach:07:0,patsahan:12:0}. 
 Neither this divergency nor the continuous transition exist in reality.
 This disadvantage is cured in the modified DFT for mixtures, where the  variance of the local volume fraction of the components is
 taken into account~\cite{ciach:11:2}. In Ref.\cite{Ciach:20:1}, it was
 shown that the results of the modified DFT agree semiquantitatively with Monte Carlo (MC) simulations for 
 the binary mixture with equal diameters of the particles of the two species. 
 In this work, we consider  binary mixtures 
with different size ratios of the components, and
 calculate correlation functions within 
 the theory developed in Ref.\cite{ciach:11:2,Ciach:20:1}. 
 
 Simulations of systems with Coulombic interactions are very difficult too, especially in the case of large size 
 asymmetry~\cite{orkoulas:99:0,yan:99:0,cheong:03:0}. 
 It is important to note, however  that  the periodic 
 arrangement of the particles of the two species is determined by the excess of the energy associated with 
 the density or concentration wave superimposed 
 on the average density or concentration, respectively. As shown in Ref.\cite{ciach:11:2}, 
 two types of local order in binary mixtures can occur - either alternating dense and dilute 
 regions are formed, or there are alternating regions rich in the first and in the second component. 
 For the local order formation, the wavenumber corresponding to the
 largest decrease of the energy compared to the homogeneous state is crucial.
 For this reason, similar structures are expected for different interaction potentials, if in 
 Fourier representation they assume extrema for the same wavenumbers~\cite{ciach:13:0}. We can  study 
 model systems that are expected to have similar local structure as 
 systems composed of charged particles. To simplify the simulation studies as much as possible, 
 in this work we consider a model of hard spheres with different diameters of the particles of the two species, 
 and assume  a square-well potential between particles of different kind. 
 With such interactions, nearest neighbours of different species are favored energetically over the nearest 
 neighbours of particles of the same species, as in 
 the case of a binary mixture of charged particles. 
 
 The objective of our study in this work is twofold. First, we want to determine how the size asymmetry influences the periodic 
 arrangement of particles, when the neighbourhood of different particles is energetically favored. Our second goal is to verify the 
 accuracy of the modified DFT for different size ratios of the particles, and for different thermodynamic states. 
 The theoretical results are compared with Monte Carlo (MC) simulations. 
 
 In sec.2, we introduce the model and describe
 briefly the simulations. In sec. 3, we summarize the theory developed in Ref.~\cite{ciach:11:2,Ciach:20:1}, and
 adopt it to the considered model and to the size-asymmetric hard-sphere reference system. In sec.4a,
 we present theoretical and simulation results for a small size asymmetry.
 Moderate and large size-asymmetry cases are described in sec.4b and 4c, respectively. The results are summarized in sec.5.

\section{The Model and the simulation method}
\subsection{The model}
We consider  the model binary mixture in which  the particles of the same species (like particles) interact through 
hard-core interactions and the particles of different species  interact through the potential $U_{12}(r)$ beyond the hard core.
In general, the particles of different species differ in their hard-sphere diameters $\sigma_{1}\ne\sigma_{2}$. 

\begin{figure}[h]
	\centering
	\includegraphics[clip,width=0.45\textwidth,angle=0]{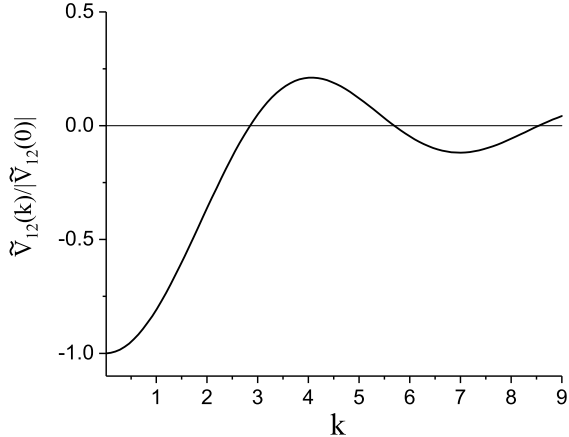}
		\vspace{-5mm}		
	\caption{\label{V12-k}
		The Fourier transform of the interaction potential between particles of different kinds 
		(Eq.~(\ref{V_sq-well}) for $a=1.2$). The wave-number $k$ is in $\sigma_{12}^{-1}$ units.
	}
\end{figure}

Thus, the pair interaction potentials are as follows: 
\begin{eqnarray}
u_{\alpha\alpha}(r) = \left\{
\begin{array}{ll}
\infty, & r<\sigma_{\alpha}\\
0,& r\geq \sigma_{\alpha}
\end{array}
\right. \,,           \qquad
u_{12}(r) = \left\{
\begin{array}{ll}
\infty, & r<\sigma_{12}\\
U_{12}(r),& r\geq \sigma_{12}
\end{array}
\right.,
\label{interact}
\end{eqnarray}
where $\sigma_{12}=(\sigma_{1}+\sigma_{2})/2$.
For $U_{12}(r)$, we choose the  square well potential, which can be presented in the form
\begin{eqnarray}
U_{12}(r)=-\varepsilon\theta(r-1)\theta(a-r),
\label{V_sq-well}
\end{eqnarray}
where $a$ and $r$ are  in $\sigma_{12}$ units. $a>1$ is the range of  the potential and $\varepsilon$ 
is the interaction strength at contact of the two unlike 
particles. 
The Fourier transform of the potential $\beta U_{12}$  has the form:
\begin{equation}
\beta\tilde{U}_{12}(k)=\frac{4\pi(ak\cos(ak)-k\cos(k)+\sin(k)-\sin(ak))}{k^3},
\label{V12_k}
\end{equation}
 where $\beta=1/(k_{B}T)$, $\beta$ is the Boltzmann constant, and the wave-number $k$ is in $\sigma_{12}^{-1}$ units. 
 The function  (\ref{V12_k}) is shown in Fig.~\ref{V12-k} for $a=1.2$.
For $a=1.2$, the first maximum  is at $k=k_{0}=4.062$.

The considered mixture undergoes two types of instability
in the MF approximation: 
the instability  (at $k=0$) connected with the gas-liquid phase separation and the instability (at $k_{0}\approx 4.062$)
connected with the appearance of local inhomogeneity at the length scale $2\pi/k_{0}$~\cite{ciach:11:2,Ciach:20:1}.

In this work, we want to compare the local structure for small, moderate and large size asymmetry, $\alpha=\sigma_{1}/\sigma_{2}$,
and choose the following values of the diameter ratio: $\alpha=0.8, 0.6, 0.25$. 
The corresponding ratios of the particle volumes, $v_{\alpha}=\pi \sigma_{\alpha}^3/6$, are $v_1/v_2=0.512, 0.216,0.015625$, respectively.

\subsection{The simulations}
The  binary mixture with $\alpha=0.8, 0.6, 0.25$
and $a=1.2\sigma_{12}$, where  $\sigma_{12}=(\sigma_{1}+\sigma_{2})/2$, has been simulated via Monte Carlo technique in the NVT ensemble.
 $N=N_{1}+N_{2}$ particles (systems I-III in Table~\ref{TableI}) are placed in a cubic box with periodical boundary conditions
 applied  in the three directions. Interactions between the particles are described in Eqs.~(\ref{interact})--(\ref{V_sq-well}). 
 The reference length for the particle size and the simulation box is $\sigma_{1}=1.0$, and the energies are $\epsilon_{1}=\epsilon_{2}=0.0$
 and  $\epsilon_{12}=-1.0$. A cut-off radius, which is the interaction range of the square well potential, depends on $\sigma_{2}$, 
 for that reason it is indicated in Table I. Each system have run $10^ {7}$ Monte Carlo steps for equilibration and 
 $10^ {6}$ for production. 
 \begin{table}[h]
 	\centering
 	\caption{\label{TableI} Simulation details for the systems considered. For systems I--II the size of the simulation box 
 		is 20$\sigma_{1}$, but for system III, the box size is 40$\sigma_{1}$. $N_{1},\zeta_{1}$ and $N_{2},\zeta_{2}$ denote
 		the number of particles and the volume fraction of the species $1$ and $2$, respectively. $R_{cut}$ 
 		is the cutoff length in $\sigma_1$ units, $\zeta_{i}=\pi\rho_{i}\sigma_{i}^3/6$, $\rho_{i}=N_{i}/V$, and  $T^*=k_BT/|\epsilon_{12}|$.}	
 	\begin{tabular}{| c | c | c | c | c | c | c | c| c | c | c} 
 		\hline
 		Systems &$\alpha$=$\sigma_{1}$/$\sigma_{2}$ &\hspace{2mm} T* \hspace{2mm}&  \hspace{2mm} $N_{1}$\hspace{2mm} & \hspace{2mm} $N_{2}$ \hspace{2mm} & \hspace{2mm} $\zeta_{1}$\hspace{2mm} & \hspace{2mm} $\zeta_{2}$ \hspace{2mm} & \hspace{2mm} $R_{cut}$ \hspace{2mm} \\ \hline 
 		\hline
 		I  & 0.8 & 0.1 & 2597 & 1799 & 0.17 & 0.23 & 1.35 \\ \hline   
 		II & 0.6 & 0.1 &3514  & 886 & 0.23 & 0.27 & 1.6\\ \hline  
 		III & 0.25 & 0.6 & 15890& 516 & 0.13 & 0.27 & 3.0\\ \hline 
 	\end{tabular}
 	\centering
 \end{table}

\section{Theory}
\subsection{Brief summary of the DFT for inhomogeneous mixtures}
In the case of size asymmetry, the number density of smaller particles can be larger than the number density of the big ones,
 but the volume occupied by them can be significantly smaller. For this reason, the theory developed in Ref.~\cite{ciach:11:2} is based
 on the local volume fractions, $\zeta_{\alpha}({\bf r})<1$, where in the binary mixture $\alpha=1,2$. We are interested in the 
 correlation functions in the disordered phase,
 \begin{eqnarray}
 \label{G}
  G_{\alpha\beta}({\bf r})=\langle \Delta\zeta_{\alpha}({\bf r}_0) \Delta\zeta_{\beta}({\bf r}+{\bf r}_0)\rangle,
 \end{eqnarray}
where $\Delta \zeta_{\alpha}({\bf r})=\zeta_{\alpha}({\bf r})-\bar\zeta_{\alpha}$, and
$\bar\zeta_{\alpha}$ is the average volume fraction  of the species $\alpha$ in the disordered phase. The matrix ${\bf G}$ with the elements
defined in (\ref{G}) satisfies the analog of the Ornstein-Zernicke equation, ${\bf G}={\bf C}^{-1}$, where the inverse correlation 
functions  $\tilde C_{\alpha\beta}({\bf r}_1,{\bf r}_2)$ (related to the direct correlation functions) are the
second functional derivatives with respect to $\zeta_{\alpha}({\bf r}_1)$ and $\zeta_{\beta}({\bf r}_2)$ of the functional 
\begin{equation}
\label{FF}
\beta F[\zeta_1,\zeta_2]=\beta\Omega_{co}[\zeta_1,\zeta_2]-
\ln\Big[\int D\phi_1  \int D\phi_2 e^{-\beta H_{fluc}}\Big],
\end{equation}
where 
\[
\label{Hfl}
H_{fluc}[\zeta_1,\zeta_2|\phi_1,\phi_2]=\Omega_{co}[\zeta_1+\phi_1,\zeta_2+\phi_2]-\Omega_{co}[\zeta_1,\zeta_2],
\]
and $\Omega_{co}[\zeta_1,\zeta_2]$ is
the grand potential with the local volume fractions constrained to have the fixed forms. We assume
\[
\label{U}
\Omega_{co}=\frac{1}{2}\int d{\bf r}_1\int d{\bf r}_2
 V_{\alpha \beta}(|{\bf r}_1-{\bf r}_2|)\zeta_{\alpha}({\bf r}_1)\zeta_{\beta}({\bf r}_2)-TS-\mu_{\alpha} N_{\alpha}, 
\]
where $V_{\alpha \beta}(r)=U_{\alpha \beta}(r)/(v_{\alpha}v_{\beta})$ is the interaction potential (in appropriate units) 
between the species $\alpha$ and $\beta$ separated by the distance $r$, 
 the entropy $S$ satisfies the relation $-TS=\int f_h(\zeta_1({\bf r}),\zeta_2({\bf r}))d{\bf r}$, 
 where $f_h$ is the free-energy density of the 
hard-core reference 
system in the local-density approximation, and $\mu_{\alpha}, N_{\alpha}$ are the chemical potential and the number of particles
of the species $\alpha$, respectively.
$\phi_{\alpha}({\bf r})$ is the local fluctuation of the volume fraction of the component $\alpha$.

In MF, the second term on the RHS of Eq.(\ref{FF}) is neglected.
In the lowest-order nontrivial approximation beyond MF~\cite{ciach:11:2},
\begin{eqnarray}
\label{C}
\tilde C_{\alpha\beta}(k)=\beta \tilde V_{\alpha\beta}(k)+A_{\alpha\beta}+\frac{A_{\alpha\beta\gamma\delta}}{2}{\cal G}_{\gamma\delta},
\end{eqnarray}
where $\tilde f(k)$ denotes the function $f$ in Fourier representation, and the summation convention is used. In the above equation,
\begin{eqnarray}
\label{A4}
A_{\alpha_1....\alpha_j}=\frac{\partial^j \beta f_h( \zeta_1, \zeta_2)}
{\partial\zeta_{\alpha_1}...\partial\zeta_{\alpha_j}},
\end{eqnarray}
with $\alpha_i=1,2$.
Note that in this approximation, the dependence of $\tilde C_{\alpha\beta}(k)$ on $k$ comes only from $\beta \tilde V_{\alpha\beta}(k)$.
The last term in Eq.(\ref{C}) is the  fluctuation contribution, 
and comes from the last term in (\ref{FF}) in the Brazovskii-type approximation~\cite{brazovskii:75:0}.
Here, ${\cal G}_{\gamma\delta}$ 
denotes the integral
\begin{eqnarray}
\label{calG}
{\cal G}_{\gamma\delta}=\int\frac{d{\bf k}}{(2\pi)^3}\tilde G_{\gamma\delta}(k).
\end{eqnarray}
Eqs.(\ref{C})-(\ref{calG}) have to be solved self-consistently. In general, it is a nontrivial task.

 We focus on the disordered inhomogeneous phase and  assume that the inhomogeneities 
occur on a well-defined length scale. In such a case, the peak of $\tilde G_{\gamma\delta}(k)$ is high and narrow. 
For functions with a high, narrow peak, the main contribution to the integral comes 
from the vicinity of the maximum.  
We assume that the maximum of all the integrands
in (\ref{calG}) is very close to the minimum  at $k=k_0$ of $\det \tilde {\bf C}(k)$,
and we make the approximation
\begin{equation}
\label{calG2}
{\cal G}_{\alpha\beta}=[ \tilde C_{\alpha\beta}(k_0)]{\cal G},
\end{equation}
where $[ \tilde C_{\alpha\alpha}(k)]=\tilde C_{\beta\beta}(k)$ and
$[ \tilde C_{\alpha\beta}(k)]=-\tilde C_{\alpha\beta}(k)$ for $\alpha\ne\beta$, and
\begin{eqnarray}
\label{calG3}
{\cal G}=\int\frac{d{\bf k}}{(2\pi)^3}\frac{1}{\det \tilde {\bf C}(k)}.
\end{eqnarray}
Near the minimum at $k_0$, we have the approximation
\begin{eqnarray}
\label{detC}
\det \tilde {\bf C}(k)= D_0+\frac{\beta\tilde W^{''}(k_0)}{2}(k-k_0)^2+...
\end{eqnarray}
where $\tilde W^{''}(k_0)$ depends on the interaction potentials, and
\begin{eqnarray}
\label{D0}
D_0
=\det\tilde {\bf C}(k_0).
\end{eqnarray}
From the approximation (\ref{detC}) and (\ref{calG3}), we obtain~\cite{ciach:11:2,ciach:12:0}
\begin{eqnarray}
\label{calGexpl}
{\cal G}\approx
\frac{k_0^2}{\pi\sqrt{2\beta \tilde W^{''}(k_0)D_0}}.
\end{eqnarray}

With all the above assumptions, the problem reduces to determination of the minimum of $\det \tilde {\bf C}(k)$,
and to a solution of  3 algebraic equations for $ \tilde C_{\alpha\beta}(k_0)$ (see Eq.(\ref{C}) for $k=k_0$), because 
\begin{eqnarray}
\label{C(k)}
\tilde C_{\alpha\beta}(k)=  \tilde C_{\alpha\beta}(k_0) +\beta(\tilde V_{\alpha\beta}(k)-\tilde V_{\alpha\beta}(k_0)).
\end{eqnarray}

\subsection{Correlation functions in the considered model}

For the model (\ref{interact})-(\ref{V_sq-well}), 
 the closed set of 4 equations for the unknowns $k_{0}$ and $\tilde C_{\alpha\beta}(k_0)$ takes the form
\begin{eqnarray}
\label{k0}
\tilde V_{12}^{'}(k_0)=0,
\end{eqnarray}
\begin{eqnarray}
&\tilde C_{11}(k_0)= A_{11}+
\frac{k_0^2}{2\pi\sqrt{2\beta \tilde W^{''}(k_0)D_0}}\left[A_{1111}\tilde C_{22}(k_0)-2A_{1112}\tilde C_{12}(k_0)+A_{1122}\tilde C_{11}(k_0)\right], 
\label{C11(k0)} \\
&\tilde C_{22}(k_0)= A_{22}+
\frac{k_0^2}{2\pi\sqrt{2\beta \tilde W^{''}(k_0)D_0}}\left[A_{1122}\tilde C_{22}(k_0)-2A_{1222}\tilde C_{12}(k_0)+A_{2222}\tilde C_{11}(k_0)\right], 
\label{C22(k0)} \\
&\tilde C_{12}(k_0)=\beta\tilde V_{12}(k_0)+ A_{12}+
\frac{k_0^2}{2\pi\sqrt{2\beta \tilde W^{''}(k_0)D_0}}\left[A_{1112}\tilde C_{22}(k_0)-2A_{1122}\tilde C_{12}(k_0)\right. \nonumber \\ 
&\left.
+A_{1222}\tilde C_{11}(k_0)\right],
\label{C12(k0)} 
\end{eqnarray}
where
\begin{eqnarray}
\tilde W^{''}(k_0)&=&-2\tilde V_{12}^{''}(k_0)\tilde C_{12}(k_0), \label{W''}
\end{eqnarray}
and we denote by $'('')$   the first (second)-order derivative of $\tilde V_{12}(k)$ with respect to the wave number $k$.  
Note that the minimum of $\det\tilde {\bf C}(k)$ corresponds to the maximum of $\tilde V_{12}(k)$.

 Once Eqs.(\ref{k0})-(\ref{C12(k0)}) are solved, the correlation functions  $\tilde C_{\alpha\beta}(k)$ can be obtained from Eq.(\ref{C(k)}).
 Finally,  the interaction potential $\tilde V_{12}(k)$ (an even function of $k$) is approximated as follows: 
 \begin{equation}
 \tilde{V}_{12}(k)\approx \tilde{V}_{12}(k_0)+\frac{V_{12}^{''}(k_0)}{8k_0^2}(k^2-k_0^2)^2.
 \label{Delta_V12} 
 \end{equation}

To calculate $\tilde G_{\alpha\beta}(k)$ in the Brazovskii-type approximation, we solve Eqs~(\ref{C11(k0)})-~(\ref{C12(k0)}) 
with respect to $\tilde C_{\alpha\beta}(k_{0})$ and from ${\bf G}={\bf C}^{-1}$, we obtain
\begin{eqnarray}
\tilde{G}_{11}(k)&=&\frac{\tilde{C}_{22}(k_{0})}{D(k)}, \qquad
\tilde{G}_{22}(k)=\frac{\tilde{C}_{11}(k_{0})}{D(k)}, \qquad
\label{G22-fl-sw} \\
\tilde{G}_{12}(k)&=&-\frac{\tilde{C}_{12}(k_{0})+\beta\Delta\tilde{V}_{12}(k)}{D(k)},
\label{G12-fl-sw}
\end{eqnarray}
where 
\begin{eqnarray}
 D(k)=\tilde{C}_{11}(k_{0})\tilde{C}_{22}(k_{0})-(\tilde{C}_{12}(k_{0})+\beta\Delta\tilde{V}_{12}(k))^{2},
 \label{D}
\end{eqnarray}
  and $\Delta\tilde{V}_{12}(k)=\tilde{V}_{12}(k)-\tilde{V}_{12}(k_{0})\approx \frac{V_{12}^{''}(k_0)}{8k_0^2}(k^2-k_0^2)^2$ 
  (see Eq.~(\ref{Delta_V12})).
  
 In MF, the correlation functions are given by Eqs.(\ref{G22-fl-sw})-(\ref{D}), but with 
 $\tilde{C}_{\alpha\alpha}(k)$ approximated by $A_{\alpha\alpha}$, and $\tilde{C}_{12}(k)$ approximated by $A_{12}+\beta\tilde V_{12}(k)$.
The correlation functions diverge for vanishing denominator in (\ref{G22-fl-sw})-(\ref{G12-fl-sw}).
From  the MF equation $D(k)=A_{11}A_{22}-(A_{12}+\beta\tilde{V}_{12}(k))^{2}=0$ one can 
get the expressions for both the gas-liquid spinodals and  the $\lambda$-surfaces:	 
\begin{eqnarray}
T_{sp}^*&=-&\left(\frac{3}{4\pi}\right)^2\frac{(1+\alpha)^6}{\alpha^3}\frac{\tilde{V}_{12}(0)}{\sqrt{A_{11}A_{22}}+A_{12}}, 
\label{GLspin_as} 
\\
T_{\lambda}^*&=&\left(\frac{3}{4\pi}\right)^2\frac{(1+\alpha)^6}{\alpha^3}\frac{\tilde{V}_{12}(k_0)}{\sqrt{A_{11}A_{22}}-A_{12}},
\label{lambda-line_as} 
\end{eqnarray}
where  $T^*=k_{B}T/\varepsilon$, and
\begin{eqnarray}
\tilde{V}_{12}(k_{0}=0)=-\frac{4}{3}\pi(a^3-1), \qquad
\tilde{V}_{12}(k_{0}=4.062, a=1.2)\simeq 0.644.
\end{eqnarray}

\subsection{Free energy for a mixture of hard-spheres with unequal diameters}
We present the free-energy density of the reference system as follows:
\begin{equation}
f_h(\zeta_1,\zeta_2)=\zeta_1\ln \zeta_1+\zeta_2\ln \zeta_2 + 
f_{ex}(\zeta_1,\zeta_2),
\label{free_energy}
\end{equation}
where  in the Carnahan-Starling approximation $f_{ex}(\zeta_{1},\zeta_{2})$
has the form~\cite{Mansoori:71}:
\begin{eqnarray}
f_{ex}(\zeta_1,\zeta_2)=\frac{6}{\pi}\left(\frac{1+\alpha}{2\alpha}\right)^{3}(\zeta_{1}+\alpha^3\zeta_{2})\left[-1-\frac{3}{2}(1-y_{1}+y_{2}+y_{3})
\right.
 \nonumber \\
\left.
+\frac{3y_{2}+2y_{3}}{1-\zeta}+\frac{3}{2}\frac{(1-y_{1}-y_{2}-y_{3}/3)}{(1-\zeta)^{2}}+(y_{3}-1)\ln(1-\zeta)\right], \label{C.1}
\end{eqnarray}
with
\begin{eqnarray}
y_{1}=\Delta_{12}\frac{1+\alpha}{\sqrt{\alpha}}, \qquad
y_{2}=\Delta_{12}\frac{\zeta_{1}+\alpha\zeta_{2}}{\sqrt{\alpha}\zeta}, \qquad
y_{3}=\frac{(\zeta_{1}+\alpha\zeta_{2})^{3}}{(\zeta_{1}+\alpha^{3}\zeta_{2})\zeta^{2}},
\label{C.2}
\end{eqnarray}
\begin{equation}
\Delta_{12}=\frac{\zeta_{1}\zeta_{2}(\alpha-1)^{2}\sqrt{\alpha}}{(\zeta_{1}+\alpha^{3}\zeta_{2})\zeta},
\label{C.3}
\end{equation}
and  $\zeta=\zeta_1+\zeta_2$ ($\zeta_{\alpha}=\pi\rho_{\alpha}\sigma_{\alpha}^3/6$).
For $\alpha=1$, $f_{ex}(\zeta_1,\zeta_2)$ reduces to the  free energy density for a one-component case $f_{ex}(\zeta)$. 

Introducing  the concentration of the $2$nd species,  $x=\rho_{2}/\rho$, we can present the packing fractions $\zeta_{1}$
and $\zeta_{2}$ as follows:
\begin{equation}
\zeta_{1}=\frac{(1-x)\alpha^{3}\zeta}{x+(1-x)\alpha^{3}}, \qquad
\zeta_{2}=\frac{x\zeta}{x+(1-x)\alpha^{3}}.
\label{zeta-i} 
\end{equation}
Alternatively, the fraction of the volume occupied by the larger particles, $c=\zeta_2/\zeta$, can be considered.
Using Eqs.~(\ref{free_energy})-(\ref{C.3}), one can get explicit expressions for $A_{\alpha\beta}$ and $A_{\alpha\beta\gamma\nu}$. 
For the fixed diameter ratio $\alpha$, they are functions of both the total packing fraction $\zeta$ and either the concentration $x$, or $c$.

\section{Results}

\subsection{Case $\alpha=0.8$} 

\subsubsection{Theoretical results} 

\begin{figure}[h]
	\centering
	\includegraphics[clip,width=0.5\textwidth,angle=0]{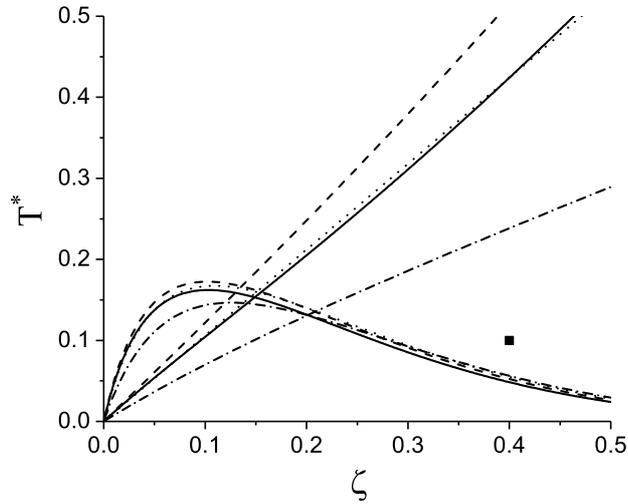}
		\vspace{-5mm}		
	\caption{\label{spinodals_as08}
		Gas-liquid spinodals and $\lambda$-lines for the model (\ref{V_sq-well}) at $a=1.2$ and $\alpha=0.8$ 
		for different concentrations. Solid lines: $c=0.33$  ($x=0.2$), dashed lines: $c=0.57$ ($x=0.4$), dotted lines: $c=0.75$ ($x=0.6$), and dash-dotted lines: $c=0.89$ ($x=0.8$).
		The filled square denotes the thermodynamic state located below the $\lambda$-surface but above the gas-liquid spinodal,
		for which the correlation functions are shown in Fig.\ref{G_k_as08-fl} (see  text for more details).
	}
\end{figure}

We start with  the case of small size asymmetry $\alpha=0.8$ ($\sigma_{2}=1.25\sigma_{1}$).
In Fig.~\ref{spinodals_as08}, we present the $T^*$-$\zeta$-plots of the MF boundaries of the stability of the disordered phase
(see Eqs.~(\ref{GLspin_as})-(\ref{lambda-line_as}))
 for a set of fixed  concentrations. In Fig.~\ref{spinodals_as08}, the curves with maxima are
the gas-liquid spinodals, while  the $\lambda$-lines are presented as straight lines. As one can see,   the dependence of $T^*$ on the
concentration at the fixed  $\zeta$ is nonmonotonic (for  both types of instability). This is in contrast to the case 
of a size-symmetric mixture \cite{Ciach:20:1}. 

\begin{figure}[h]
	\centering
	 	\includegraphics[clip,width=0.42\textwidth,angle=0]{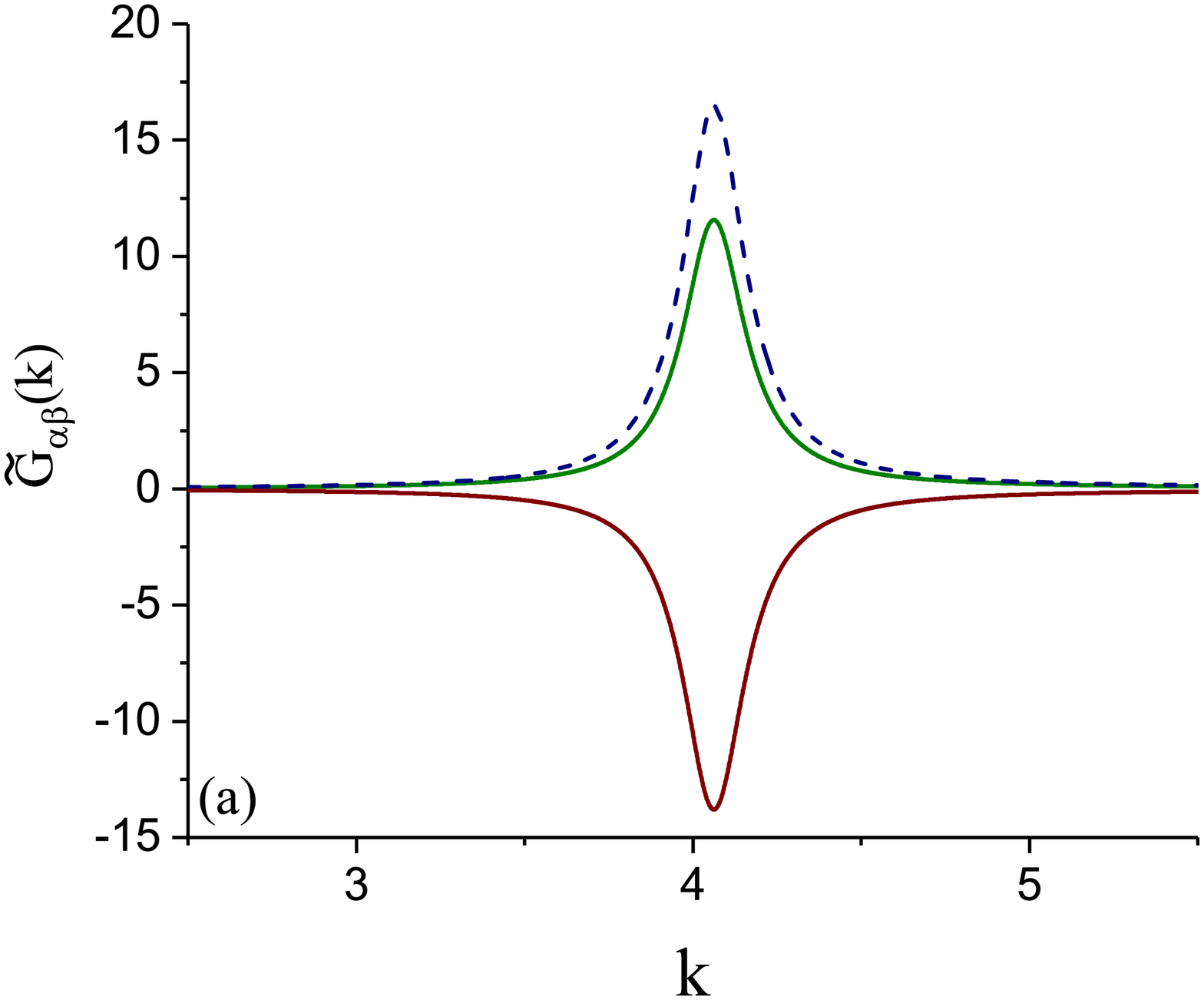}
	 	\qquad
	\includegraphics[clip,width=0.42\textwidth,angle=0]{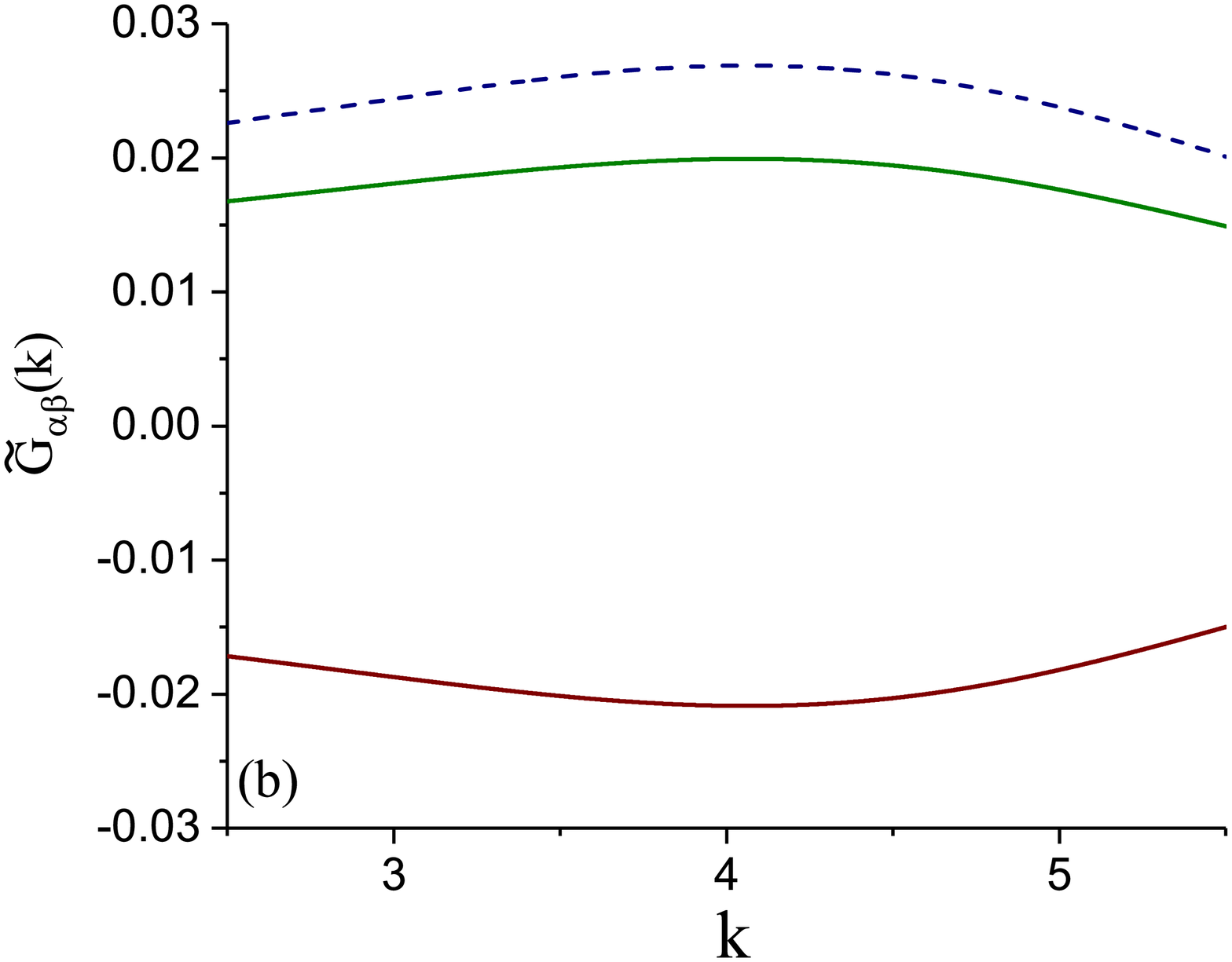}
	\caption{\label{GIJ_as08-mf_k} (Colour online)
		Correlation functions $\tilde G_{\alpha\beta}(k)$  for $\alpha=0.8$ in MF approximation (panel a) and with the effect of fluctuations 
		taken into account (panel b) for the thermodynamic state  above the $\lambda$-surface:  $T^*=0.52$, $\zeta=0.4$ and $c=0.57$ ($x=0.4$). Upper solid lines: $\tilde G_{11}(k)$,   dashed lines: $\tilde G_{22}(k)$, 
		and  lower solid lines: $\tilde G_{12}(k)$. 
	}
\end{figure}

First, we calculate the MF correlation functions in Fourier representation
above the $\lambda$-surface. The corresponding correlation functions 
$\tilde{G}_{\alpha\beta}(k)$ are shown in Fig.~\ref{GIJ_as08-mf_k} (panel a) for $T^{*}=0.52$, $\zeta=0.4$ and $c=0.57$. 
The maximum (minimum) of $\tilde{G}_{\alpha\beta}(k)$ corresponds to the maximum of the interaction potential $\tilde V_{12}(k)$ and it is located at $k=k_{0}$ (see Fig.~\ref{V12-k}).

\begin{figure}[h]
	\centering
	\includegraphics[clip,width=0.42\textwidth,angle=0]{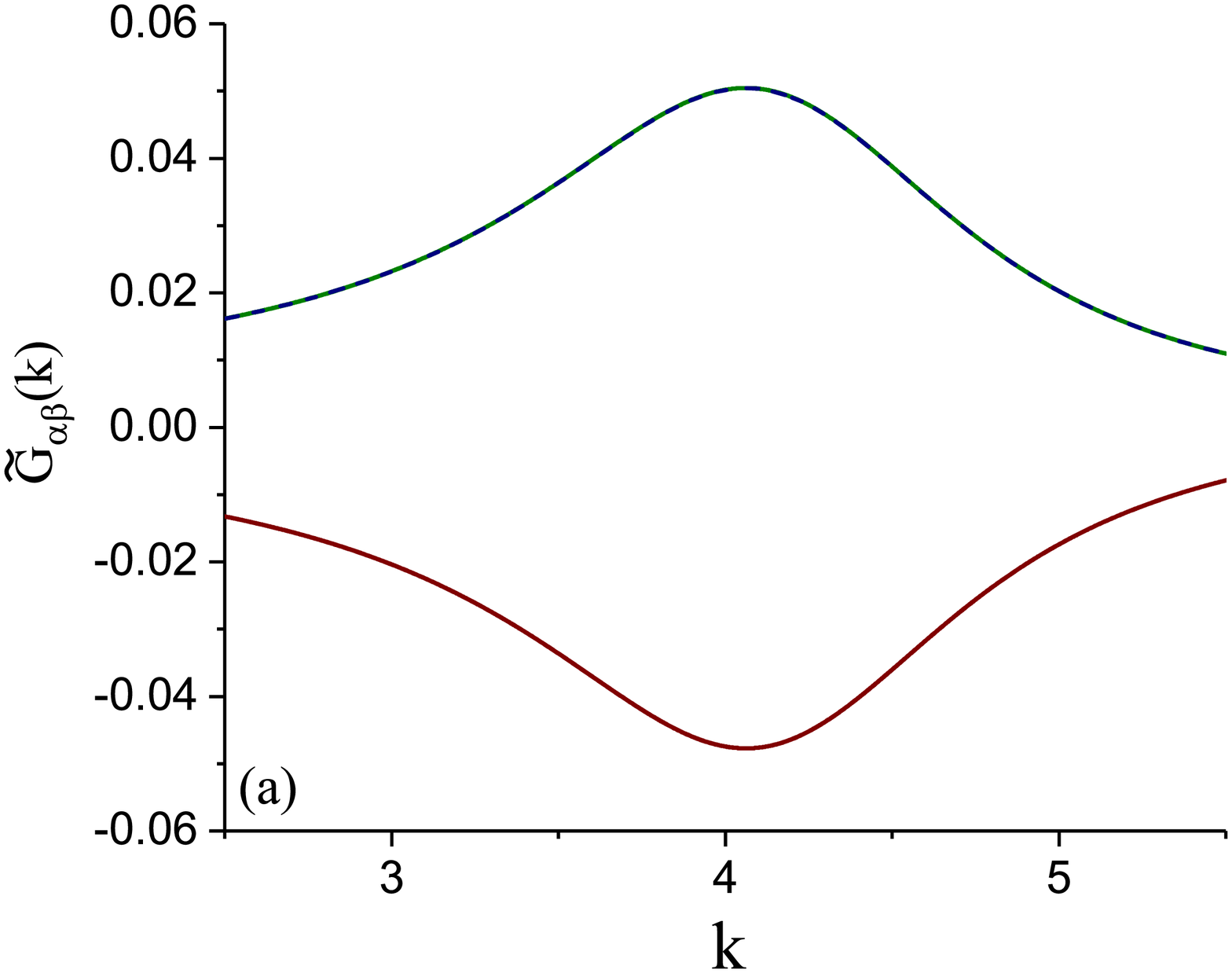} 
	\qquad
	\includegraphics[clip,width=0.42\textwidth,angle=0]{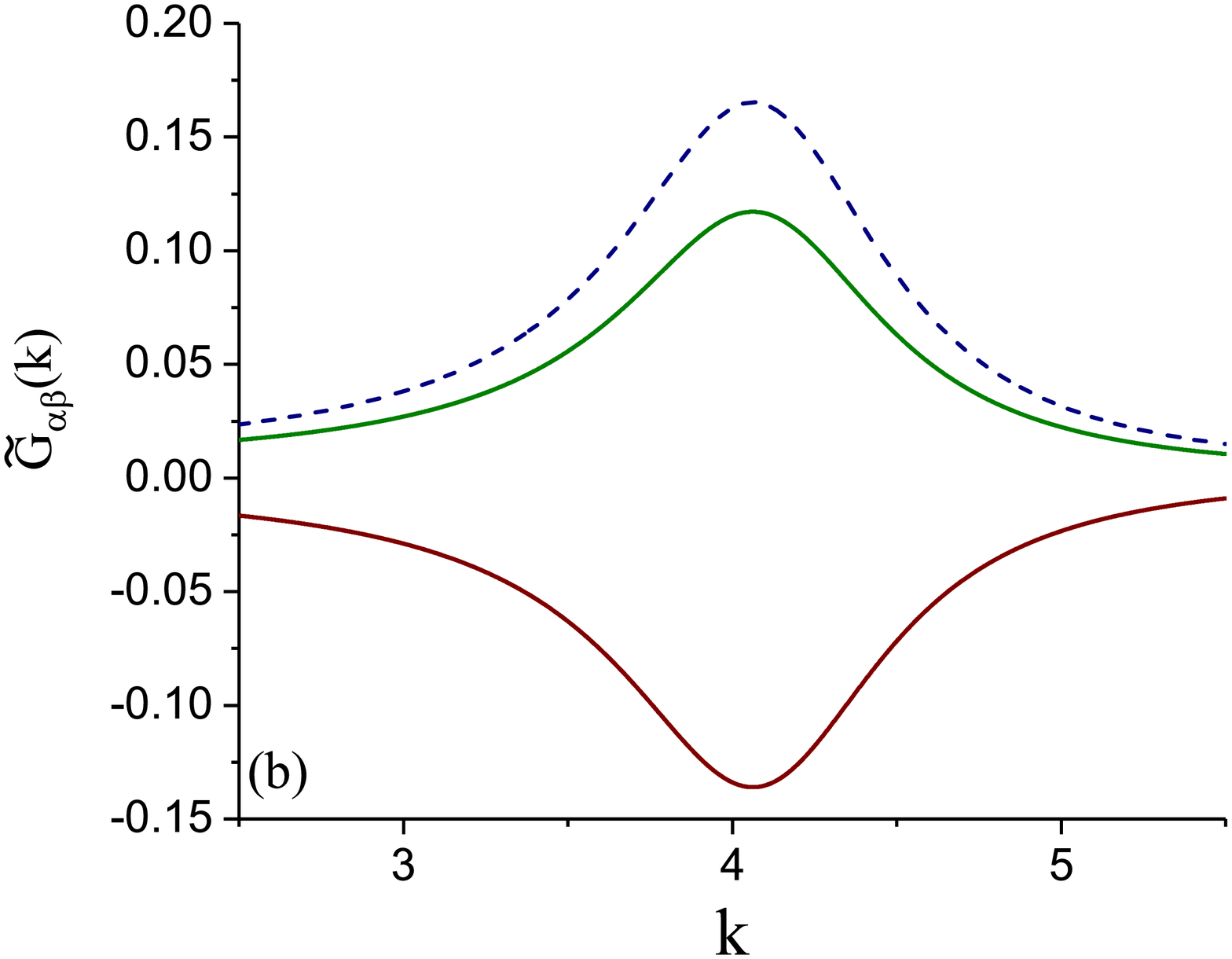} 
	\caption{\label{G_k_as08-fl} (Colour online)
		Correlation functions in Fourier representation  for $\alpha=0.8$  
		with the effect of fluctuations taken into account. Upper solid lines: $\tilde G_{11}(k)$, 
		dashed lines: $\tilde G_{22}(k)$ and lower solid lines: $\tilde G_{12}(k)$.    $T^*=0.1$, $\zeta=0.4$  and $c=0.33$ (panel a), $c=0.57$  (panel b).
	}
\end{figure}

Now, we  go beyond MF and calculate  the correlation functions in Fourier representation,  taking into account the effect of fluctuations. We fix the total volume fraction at  $\zeta=0.4$ and  consider the temperatures above and below  the $\lambda$-surface. In Fig.~\ref{GIJ_as08-mf_k} (panel b), $\tilde G_{\alpha\beta}(k)$  are compared with the MF result for the same values of  temperature,  total volume fraction, and concentration. It is seen that the maxima (minimum) of the correlation functions for the temperature above the $\lambda$-surface become very  flat when the fluctuations are taken into account and $k_{0}$  is not shifted in this case.
In Fig~\ref{G_k_as08-fl},  we present the results for $\tilde G_{\alpha\beta}(k)$  for the  temperature 
$T^{*}=0.1$, the total volume fraction $\zeta=0.4$ and  for two  values of the bigger particle concentration: $c=0.33$  (panel a) and
$c=0.57$  (panel b). These thermodynamic states are located below the $\lambda$-surface and beyond 
the MF gas-liquid spinodal (see the state denoted by the filled square in Fig.~\ref{spinodals_as08}).  It should be noted that the correlation functions  $\tilde G_{\alpha\beta}(k)$ at $k_{0}$ 
take their maximal (minimal) values  for the concentration $c\approx 0.57$. For $c=0.33$,  $\tilde G_{11}(k)$ and $\tilde G_{22}(k)$ 
coincide; the maximum of $\tilde G_{11}(k)$ becomes larger (smaller) than the maximum of $\tilde G_{22}(k_{0})$ for $c<0.33$ ($c>0.33$).

\begin{figure}[h]
	\includegraphics[clip,width=0.6\textwidth,angle=0]{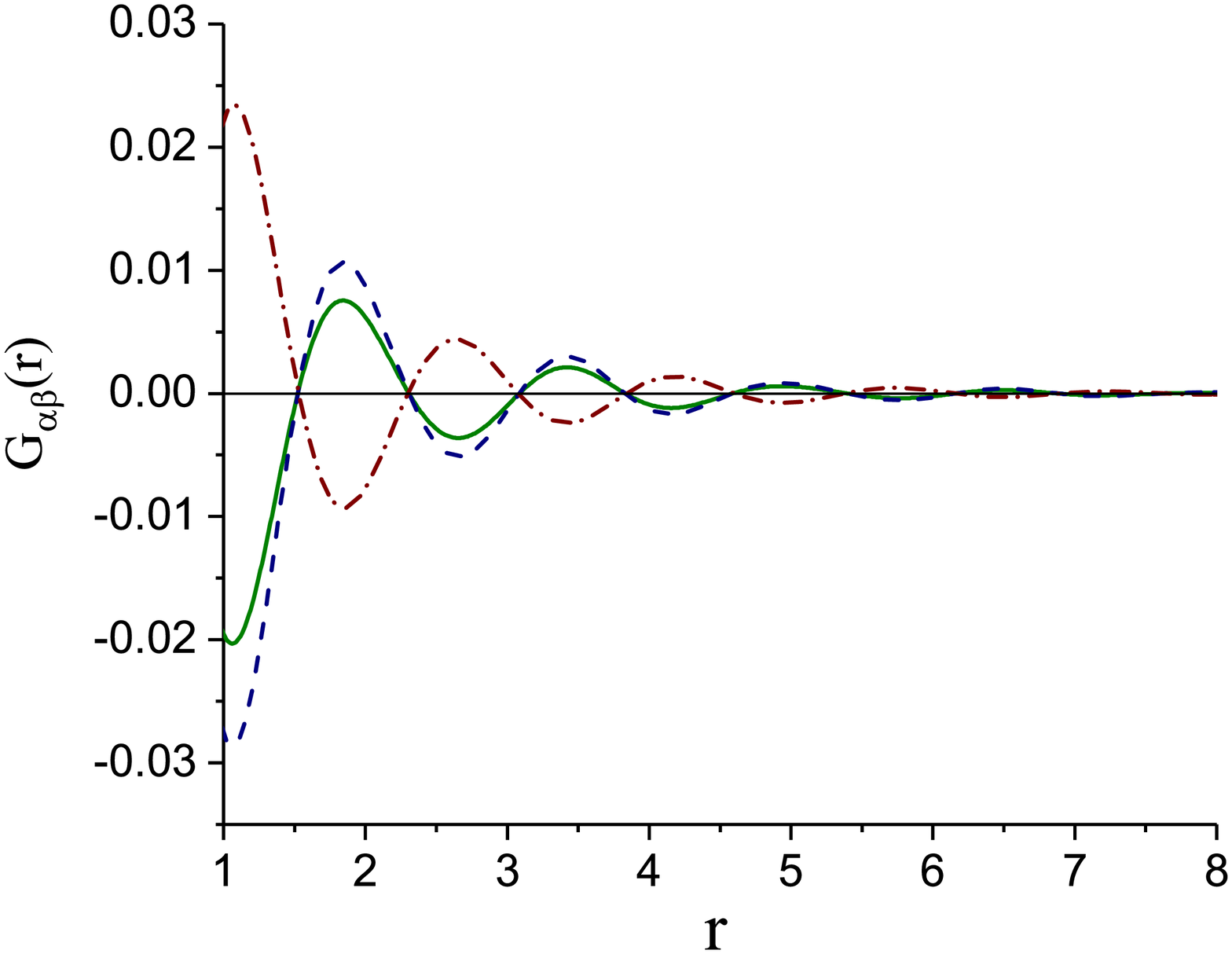}
	\caption{\label{GIJ_r_as08}
	(Colour online)	Case $\alpha=0.8$. Correlation functions in real space  for $T^*=0.1$,  $\zeta=0.4$ and   $c=0.57$ with the effect of fluctuations taken into account.
		$G_{11}(r)$ (solid line), $G_{22}(r)$ (dashed line) and $G_{12}(r)$ (dash-dotted line). $r$ is in $\sigma_{12}$ units.  
	}
\end{figure}

\begin{table}
	\centering
	\caption{\label{Table2}  The decay length $\alpha_{0}^{-1}$ and the period of oscillations $\lambda=2\pi/\alpha_{1}$ of 
	the pair correlation functions $G_{\alpha\beta}(r)$ (Eq.~(\ref{Gr})) depending on the 
	total number density $\zeta$ for  fixed values of the size asymmetry ratio $\alpha$, temperature $T^{*}$ and concentration $c$. 
	$T^{*}=k_BT/|\epsilon_{12}|$, $\alpha_{0}$ and $\alpha_{1}$ are in $\sigma_{12}^{-1}$ units}	
	\begin{tabular}{| c | c | c | c | c | c | c | c| c } 
		\hline
		 $\alpha=\sigma_{1}/\sigma_{2}$ & \hspace{2mm} $T^{*}$\hspace{2mm} &  $c=\zeta_{2}/\zeta$ &\hspace{2mm}  $\zeta$ \hspace{2mm} & \hspace{4mm} $\alpha_{0}$ \hspace{3mm}  &\hspace{3mm}  $\alpha_{1}$ \hspace{3mm} &  \hspace{3mm} $\alpha_{0}^{-1}$ \hspace{2mm} & \hspace{3mm}  $2\pi/\alpha_{1}$ \hspace{2mm} \\ \hline 
		\hline
		 0.8 & 0.1 & 0.57 &0.4  &0.4891 &4.092  & 2.044 & 1.536 \\ \hline 
		 0.8 & 0.1 & 0.57 & 0.45&0.359 & 4.078 & 2.785 &1.541 \\ \hline   
		 0.8 & 0.1 & 0.57 & 0.5 &0.267 & 4.071 & 3.750 &1.543 \\ \hline  \hline     
		 0.6 & 0.1 & 0.54 & 0.4 &0.376 &4.080  &2.657 & 1.540\\ \hline  
		 0.6 & 0.1 & 0.54 &0.45 &0.265 &4.071  & 3.775 &1.543 \\ \hline  
		 0.6 & 0.1 & 0.54 & 0.5 & 0.189 & 4.067 &5.290 &1.545 \\ \hline  \hline
		 0.25 & 0.6 & 0.67& 0.3 & 1.306& 4.267&0.766 & 1.472\\ \hline 
		 0.25 & 0.6 &0.67 &0.35 &0.676  & 4.118 & 1.478& 1.526\\ \hline 
		 0.25 & 0.6 &0.67 & 0.4 & 0.112 & 4.064 & 8.928& 1.546\\ \hline 
		   \hline		
	\end{tabular}
	\centering
\end{table}
The correlation functions  in real-space representation, obtained by
 the inverse Fourier transformation of $\tilde G_{\alpha\beta}(k)$,  are shown in Fig.\ref{GIJ_r_as08} for  $T^*=0.1$, 
 $\zeta=0.4$ and   $c=0.57$. 
 It is seen from Fig.~\ref{GIJ_r_as08} that   $G_{\alpha\beta}(r)$ show exponentially damped oscillatory behavior with 
 the period of damped oscillations $\lambda/\sigma_{12}\simeq 1.54$ which is close to $2\pi/k_{0}\simeq 1.547$. 
 In addition, $G_{12}(r)$ and $G_{\alpha\alpha}(r)$ exhibit their extrema at the same values of $r$ and the maximum of 
 $G_{\alpha\alpha}(r)$ coincides with the minimum of $G_{12}(r)$. 
 In general, $G_{\alpha\beta}(r)$ for $r\gg 1$ are described by the functions~\cite{leote:94:0}
\begin{equation}
\label{Gr}
G_{\alpha\beta}(r)=A_{\alpha\beta}e^{-\alpha_0 r}\sin(\alpha_1 r)/r.
\end{equation}
 In (\ref{Gr}),  $\alpha_{0}$ and  $\alpha_{1}$ are the imaginary and  real  parts of the leading order pole of 
$\tilde G_{\alpha\beta}(k)$ which is determined as the complex root  $k=\alpha_1\pm i\alpha_0$  of  the equation $D(k)=0$ 
(see Eq.~(\ref{D})) having the smallest imaginary part. All the $\tilde G_{\alpha\beta}(k)$ have the same pole structure and 
the same $\alpha_{1}$ and $\alpha_{0}$.  For the above-mentioned thermodynamic state,  we get
 $\alpha_{0}\sigma_{12}\simeq 0.49$, $\alpha_{1}\sigma_{12}\simeq 4.1$, $A_{11}\simeq 0.038$, $A_{22}\simeq 0.0525$, 
 and $A_{12}\simeq -0.045$.  It should be noted that the amplitudes $A_{\alpha\beta}$ satisfy the rule
 \begin{equation}
 A_{12}^2=A_{11}A_{22},
 \label{AIJ}
 \end{equation}
derived from general considerations  \cite{evans:94:0}.
\begin{figure}[h]
	\centering
	\includegraphics[clip,width=0.42\textwidth,angle=0]{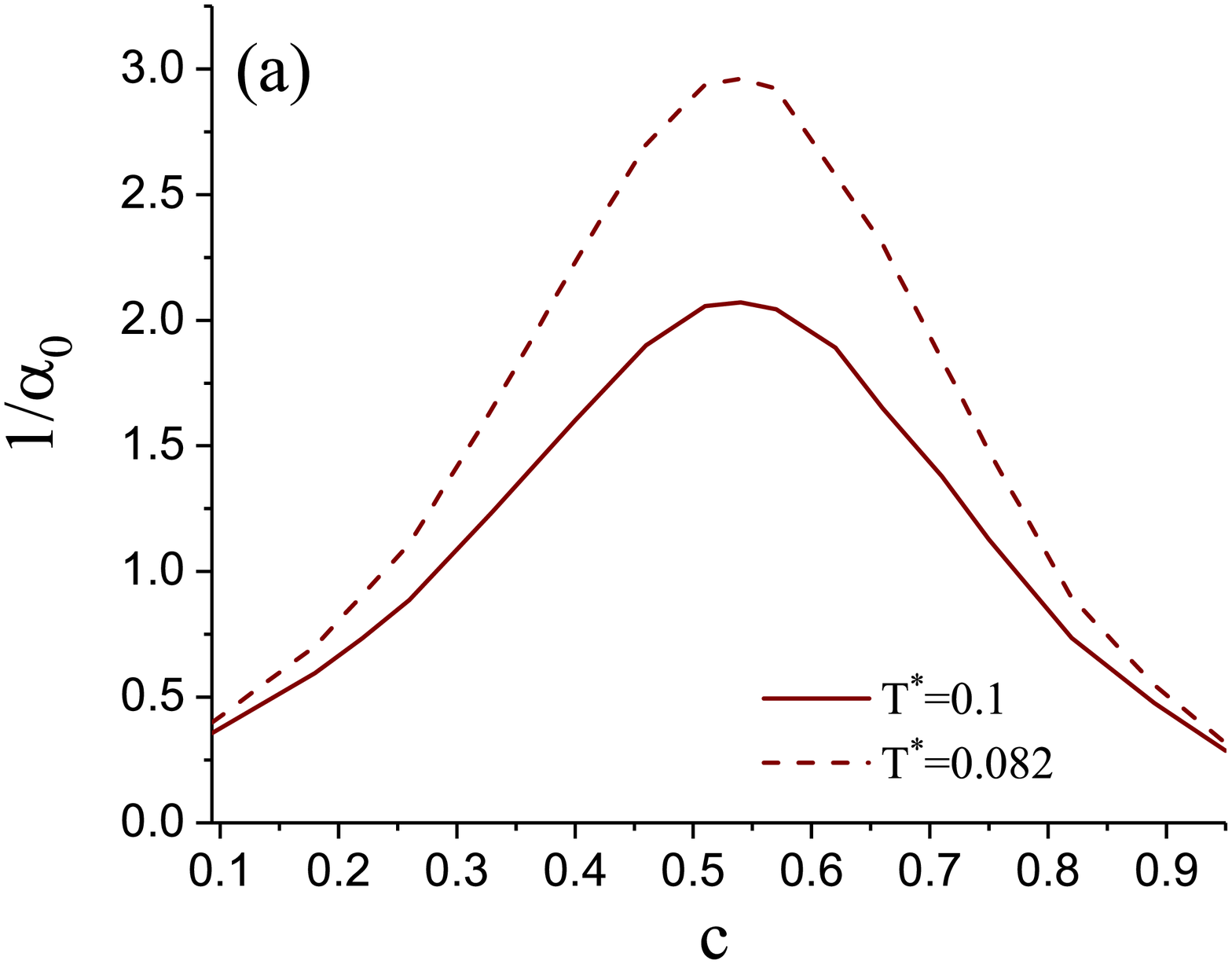}
	\qquad
	\includegraphics[clip,width=0.42\textwidth,angle=0]{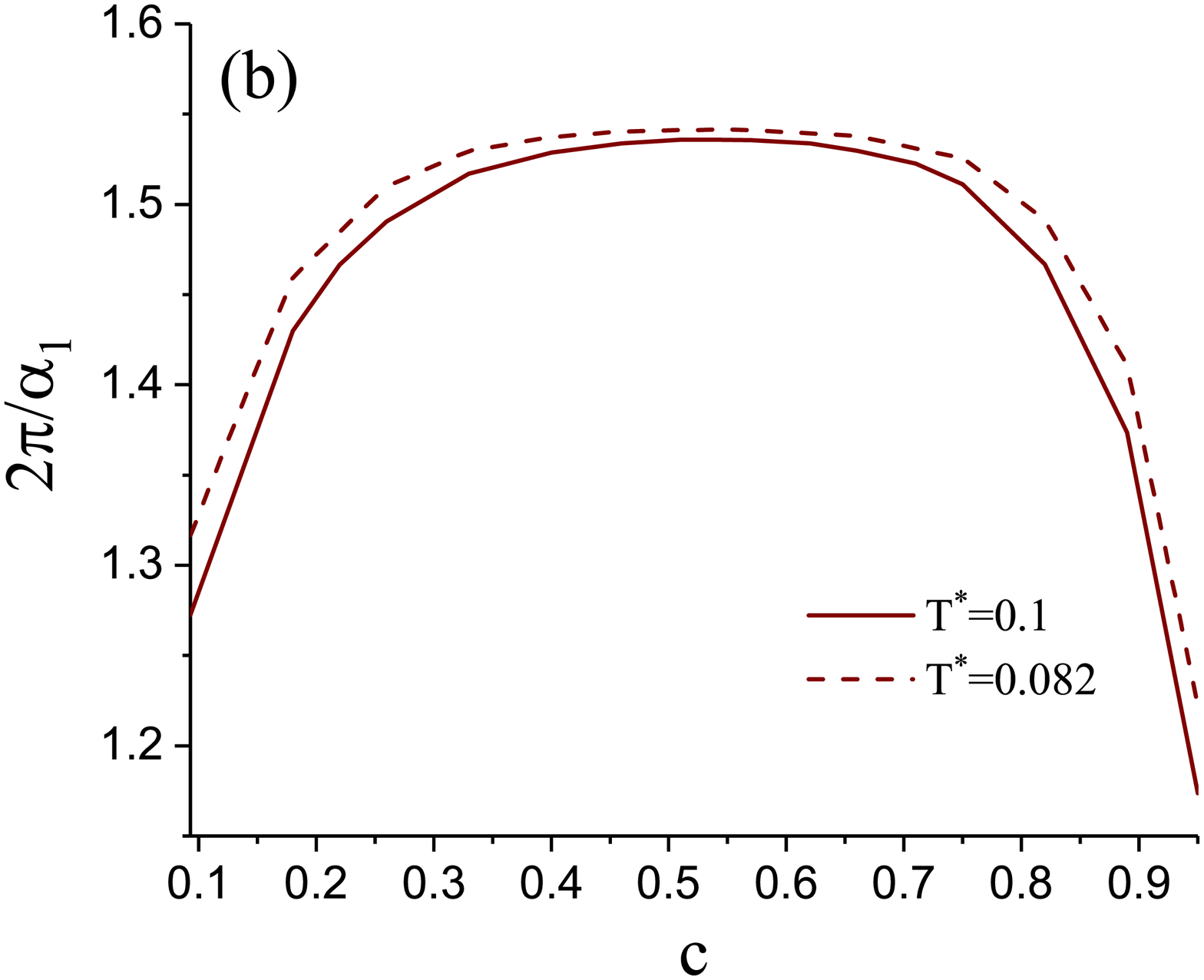}
	\caption{\label{xi-l}
	The decay length $\alpha_{0}^{-1}$ (panel~a) and the period of oscillations $\lambda=2\pi/\alpha_{1}$ 
	(panel~b) of the correlation functions $G_{\alpha\beta}$ for $\alpha=0.8$ as  
	functions of the concentration  for two values of temperature:  $T^{*}=0.082$ (dashed line)  and $T^{*}=0.1$ (solid line).  $\alpha_{0}$ and $\alpha_{1}$ are in units of $\sigma_{12}^{-1}$. 
	}
\end{figure}

The decay length,   $\alpha_{0}^{-1}$, and the period of oscillations,
$\lambda=2\pi/\alpha_{1}$, are shown in  Fig.~\ref{xi-l} as functions of the concentration $c$ 
for the total volume fraction $\zeta=0.4$ and for two values
of temperature, $T^{*}=0.082$  and $0.1$. As it is seen, both $\alpha_{0}^{-1}$ and  $\lambda$  exhibit a nonmonotonic behavior 
with the concentration: $\alpha_{0}^{-1}$ has  a pronounced maximum for $c\simeq 0.57$, whereas   $\lambda$ exhibits a wide, 
flat maximum region.  The decay length increases noticeably when  the temperature decreases,
while the temperature dependence of  $\lambda$  is very weak.  Similarly, $\alpha_{0}^{-1}$ varies quite rapidly 
on increasing (decreasing) the total volume fraction,  whereas $\lambda$  changes only very slightly (see Table~\ref{Table2}).

\subsubsection{Simulation results}
 The details of simulations of a binary mixture with $\alpha=0.8$ are given in Table~\ref{TableI} (see System I). 

\begin{figure}[h]
	\centering
	\includegraphics[clip,width=0.48\textwidth,angle=0]{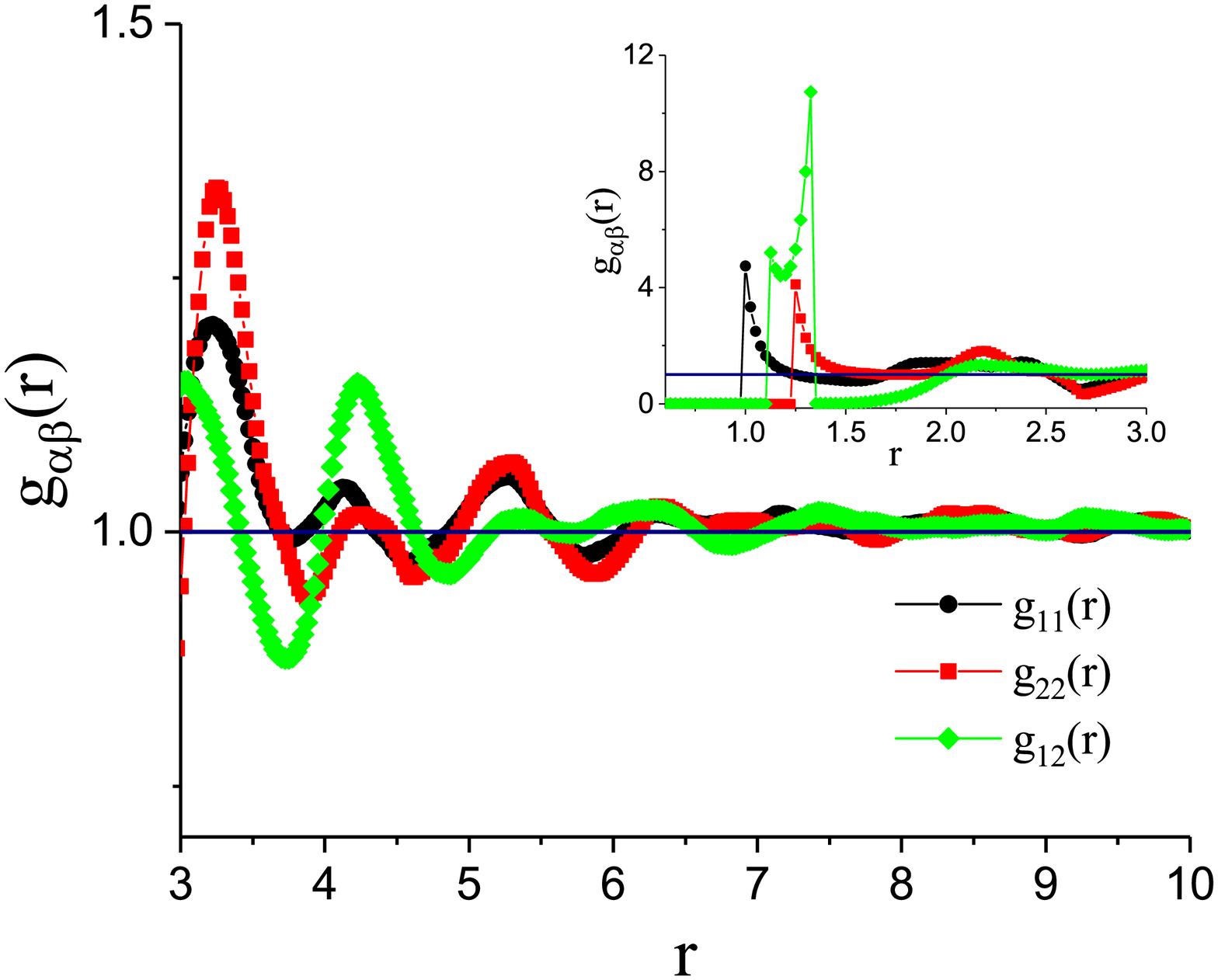}
	\quad
	\includegraphics[clip,width=0.35\textwidth,angle=0]{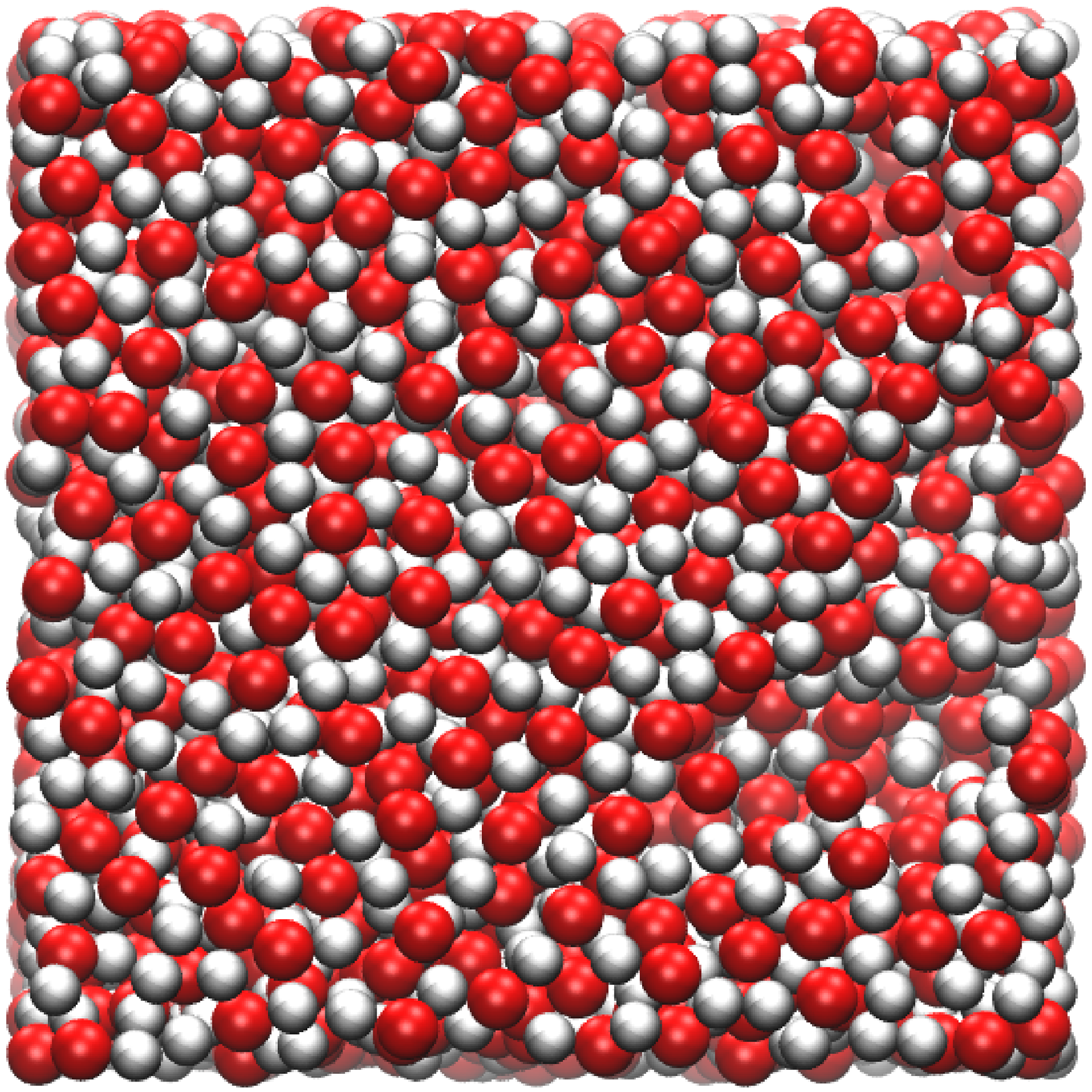}
	\caption{(Colour online) Pair distribution functions and a representative configuration  of the model with $\alpha=0.8$ for $T^*=0.1$,  total volume fraction
		$\zeta=0.4$ and concentration $c=0.57$. $r$ is in $\sigma_{1}$ units.}
	\label{sim_sqw_as08-1}
\end{figure}

In Fig.~\ref{sim_sqw_as08-1}, the pair distribution functions $g_{\alpha\beta}=\frac{G_{\alpha\beta}}{\zeta_{\alpha}\zeta_{\beta}}+1$ and a representative configuration are presented  for $T^*=0.1$, $\zeta=0.4$ and $c=0.57$. 
On the snapshot, darker (red) colour is used to denote bigger particles and lighter (cyan) colour is used for smaller particles. 
One can see that  smaller and bigger particles are located next to each other.
\begin{figure}[h]
	\centering
	\includegraphics[clip,width=0.45\textwidth,angle=0]{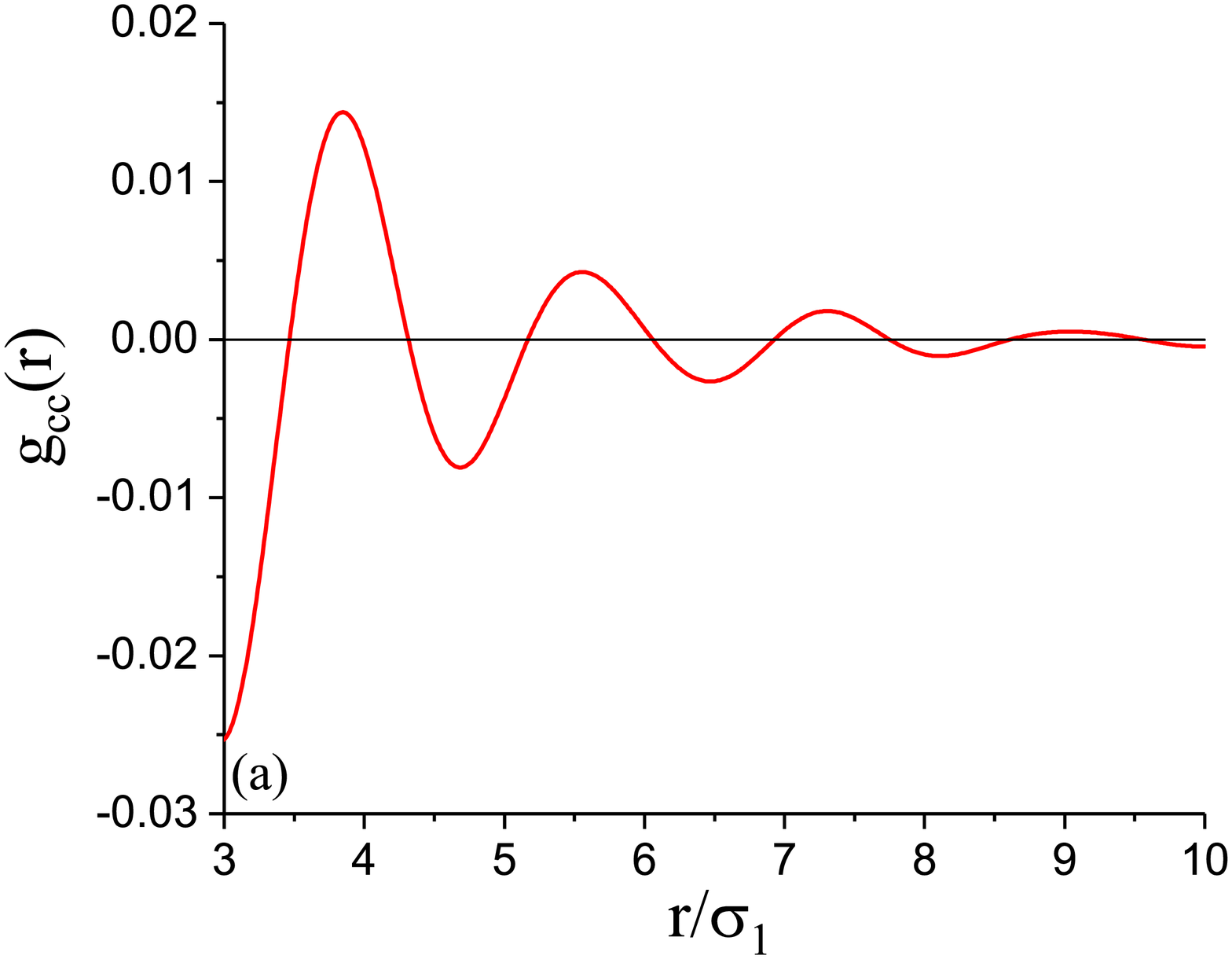} \quad
	\includegraphics[clip,width=0.45\textwidth,angle=0]{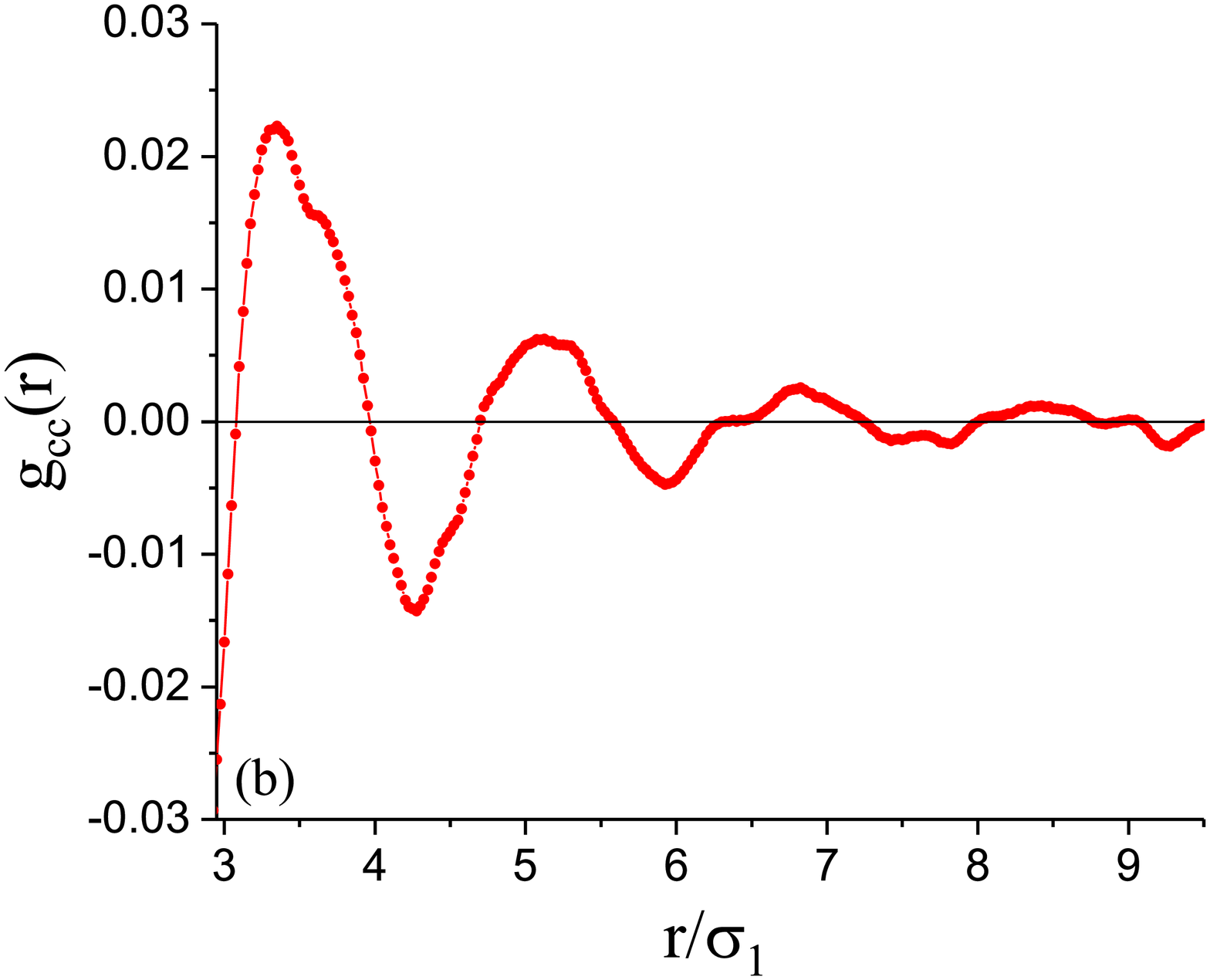}
	\caption{\label{g_cc_r_as08}   Concentration-concentration distribution function $g_{cc}(r)$ for $\alpha=0.8$.  Panel~a: theoretical results and panel~b: results of simulations. $T^*=0.1$,
		$\zeta=0.4$ and   $c=0.57$. $r$ is in $\sigma_{1}$ units.
	}
\end{figure}

For the pair distribution functions $g_{\alpha\beta}(r)$, an oscillatory decay with the period of the damped oscillations $\lambda\approx 1.1\sigma_{1}\simeq\sigma_{12}$ can be seen (Fig.~\ref{sim_sqw_as08-1}). Simultaneously,   the minima of $g_{12}(r)$ more or less coincide with the maxima of $g_{\alpha\alpha}(r)$ only for large distances. This deviations from the theoretical predictions can be related to the effect of the hard sphere packing which becomes important for the large density.
In order to separate this effect, we calculate the concentration-concentration distribution functions \cite{March1976}
\[
g_{cc}(r)=x^{2}(1-x)^{2}\left[g_{11}(r) +g_{22}(r)-2g_{12}(r) \right].
\]

In Fig.~\ref{g_cc_r_as08}, we present  $g_{cc}(r)$ obtained from the theory and from the simulations for $T^*=0.1$, $\zeta=0.4$ and  $c=0.57$. It is seen that the theoretical results are in good agreement with the simulation findings. The theory and simulations show oscillatory decay  and  the periods of damped oscillations which are close to each other. The maxima and minima of the oscillations occur for  similar $r$. 
\begin{figure}[h]
	\includegraphics[clip,width=0.33\textwidth,angle=0]{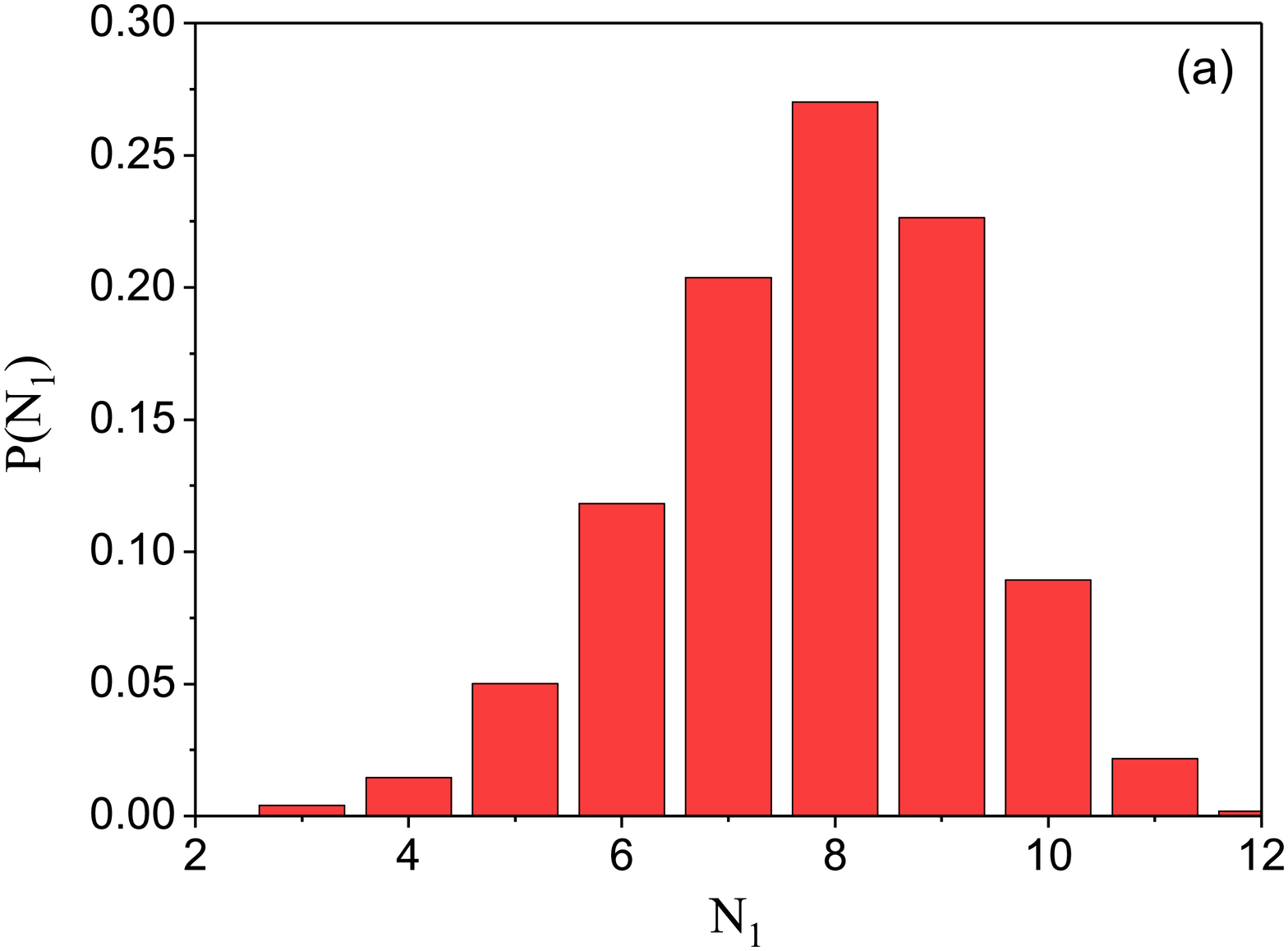}
	\includegraphics[clip,width=0.32\textwidth,angle=0]{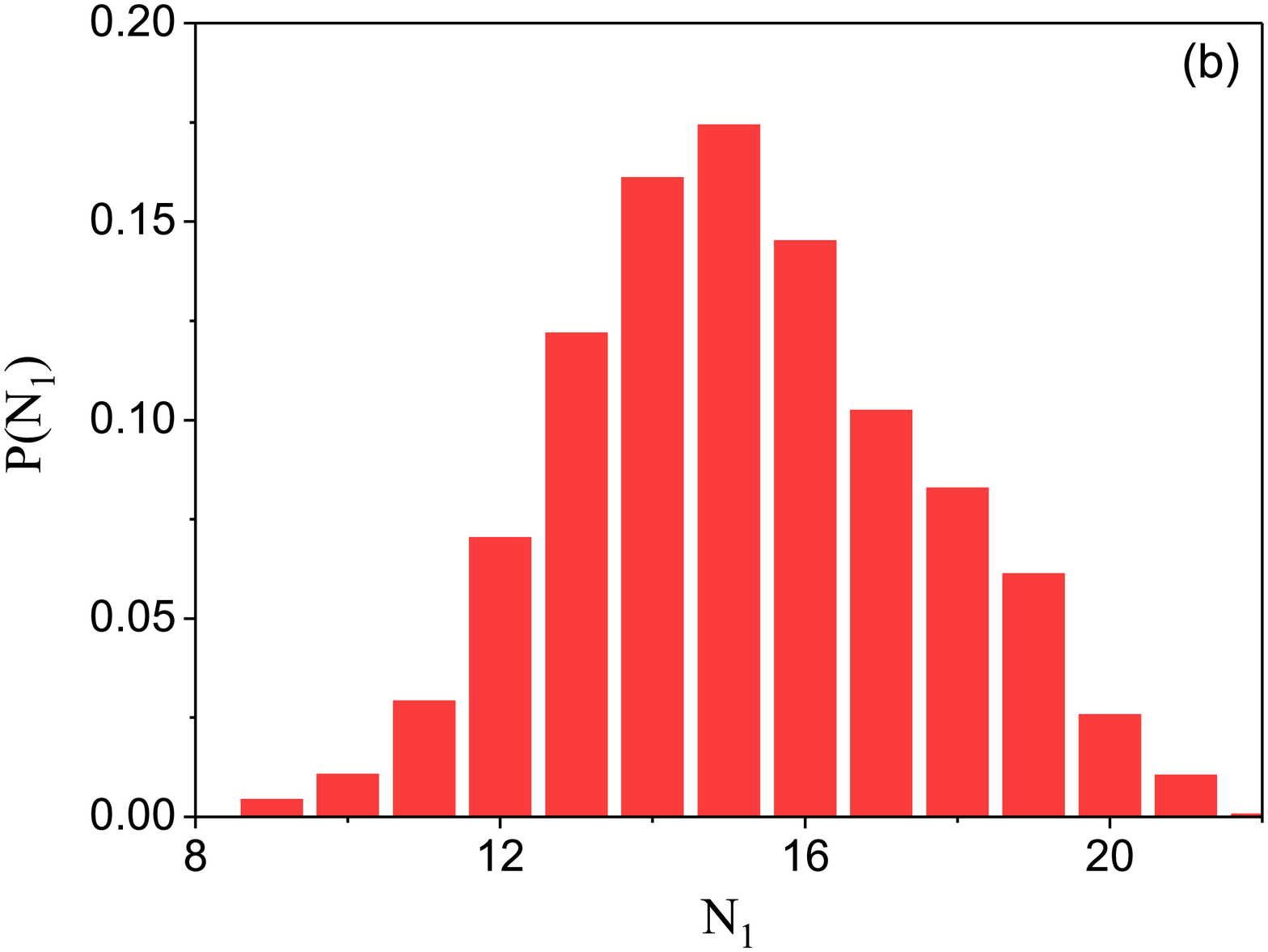}	
	\includegraphics[clip,width=0.32\textwidth,angle=0]{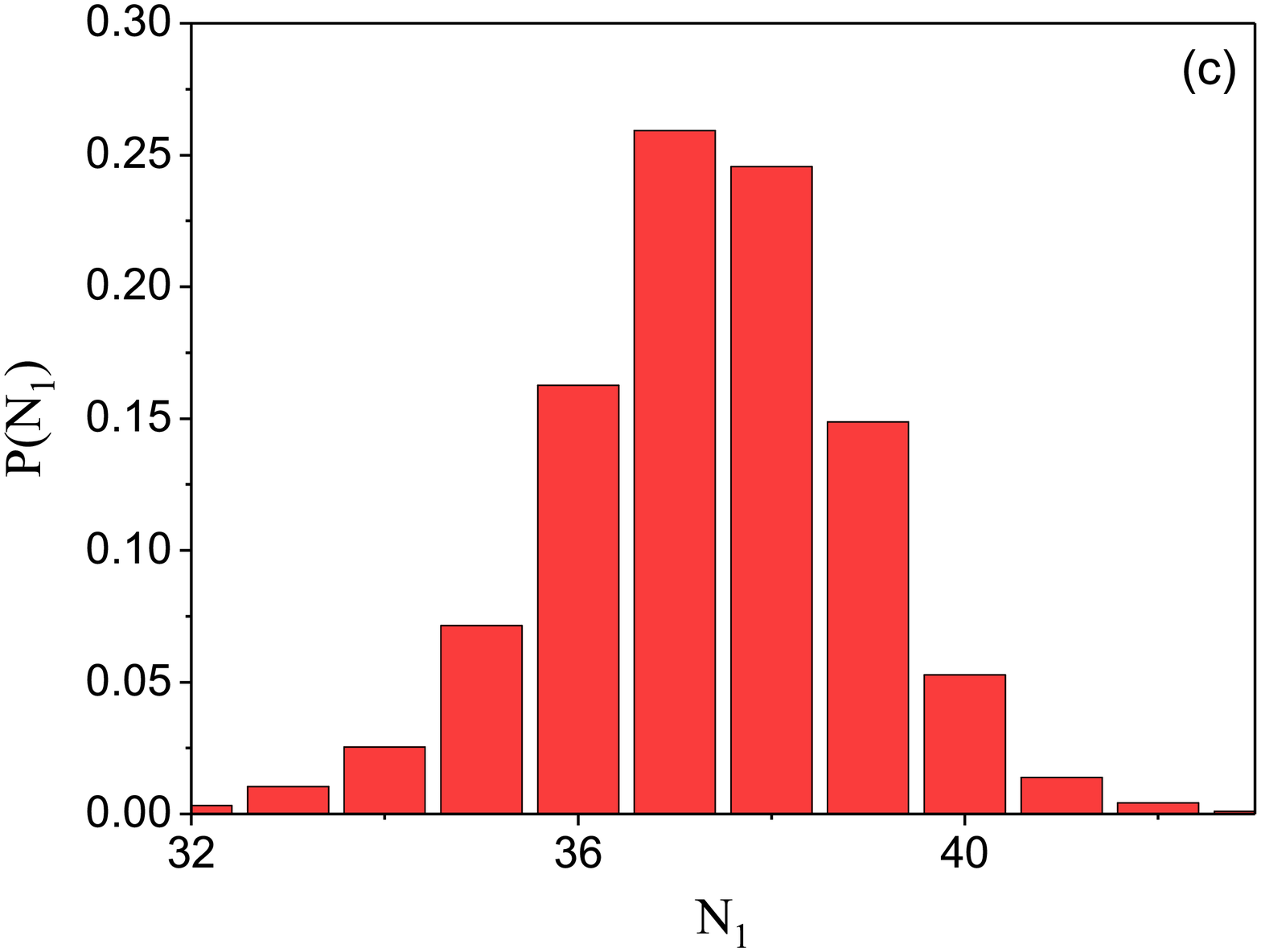}
	\caption{\label{hist_as}
		Histograms of the probability to find $N_1$ smaller particles at a distance $\sigma_{12}<r<1.2\sigma_{12}$  from the center of a bigger particle. Panel~a: $\alpha=0.8$  ($T^*=0.1$, $\zeta=0.4$, $c=0.57$), panel~b: $\alpha=0.6$ ($T^*=0.1$, $\zeta=0.5$, $c=0.54$) and panel~c: 
	 $\alpha=0.25$ ($T^*=0.6$,  $\zeta=0.4$,  $c=0.67$).
	}
\end{figure}


The theory summarized in sec. 3a can be valid only for $r\gg 1$, i.e. it allows to study ordering on the mesoscopic length scale.
It is interesting, however, to analyze the local ordering of the smaller particles in the attractive shell of the bigger one.
The degree of order is not only described by the pair distribution function that tells about the average number of the particles 
at a given distance  from the considered one, but also by the fluctuation of this number. This information can be inferred from 
simulations by determining the probability distribution of finding $N_1$ smaller particles at a distance $\sigma_{12}<r<1.2\sigma_{12}$  
from the center of the bigger one (corresponding to the attractive shell). 

The  histograms $P(N_{1})$ of the probability of finding $N_1$ smaller particles at a distance $\sigma_{12}<r<1.2\sigma_{12}$  
from the center of a bigger particle are presented  in Fig.~\ref{hist_as}. 
The results for  $\alpha=0.8$  (panel~a) indicate that the bigger particle is most likely to have 8 nearest-neighbours. 
This implies that the periodic ordering resembling  ionic crystal structure can  occur on the mesoscopic length scale.
However, fluctuations of 
$N_1$ in the attractive shell are quite large.

\subsection{Case $\alpha=0.6$} 
\subsubsection{Theoretical results} 
\begin{figure}[h]
	\centering
	\includegraphics[clip,width=0.5\textwidth,angle=0]{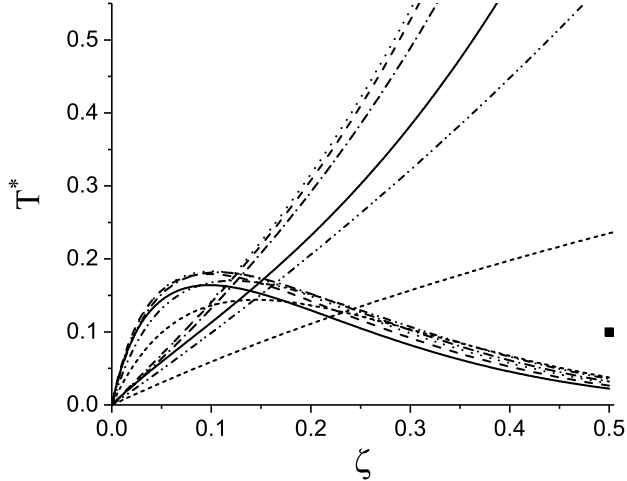}
		\vspace{-5mm}		
	\caption{\label{spinodals_as06}
		Gas-liquid spinodals and $\lambda$-lines for the model (\ref{V_sq-well}) at $a=1.2$ and $\alpha=0.6$ 
		for different  concentrations. Solid lines:  $c=0.34$ ( $x=0.1$), dashed lines: $c=0.54$  ($x=0.2$), dotted lines: $c=0.67$  ($x=0.3$), dash-dotted lines:  $c=0.76$  ($x=0.4$), dash-dot-dotted lines: $c=0.87$ ($x=0.6$), and short-dashed lines: $c=0.95$ ($x=0.8$). The filled square denotes the thermodynamic state located below the $\lambda$-surface but above the gas-liquid spinodal (see  text for details).
	}
\end{figure}

Now we consider the case of a moderate size-asymmetry,  $\alpha=0.6$ ($\sigma_{2}\simeq 1.7\sigma_{1}$).  For this mixture, the MF 
boundaries of stability  
 for the fixed values of concentration are presented in Fig.~\ref{spinodals_as06}. As for $\alpha=0.8$,  
 the dependence of $T^*$ on  the bigger particle concentration for fixed  $\zeta$ is nonmonotonic. Now, the $\lambda$-lines 
 are no longer straight lines for lower values of concentration.

\begin{figure}[h]
	\centering
	\includegraphics[clip,width=0.42\textwidth,angle=0]{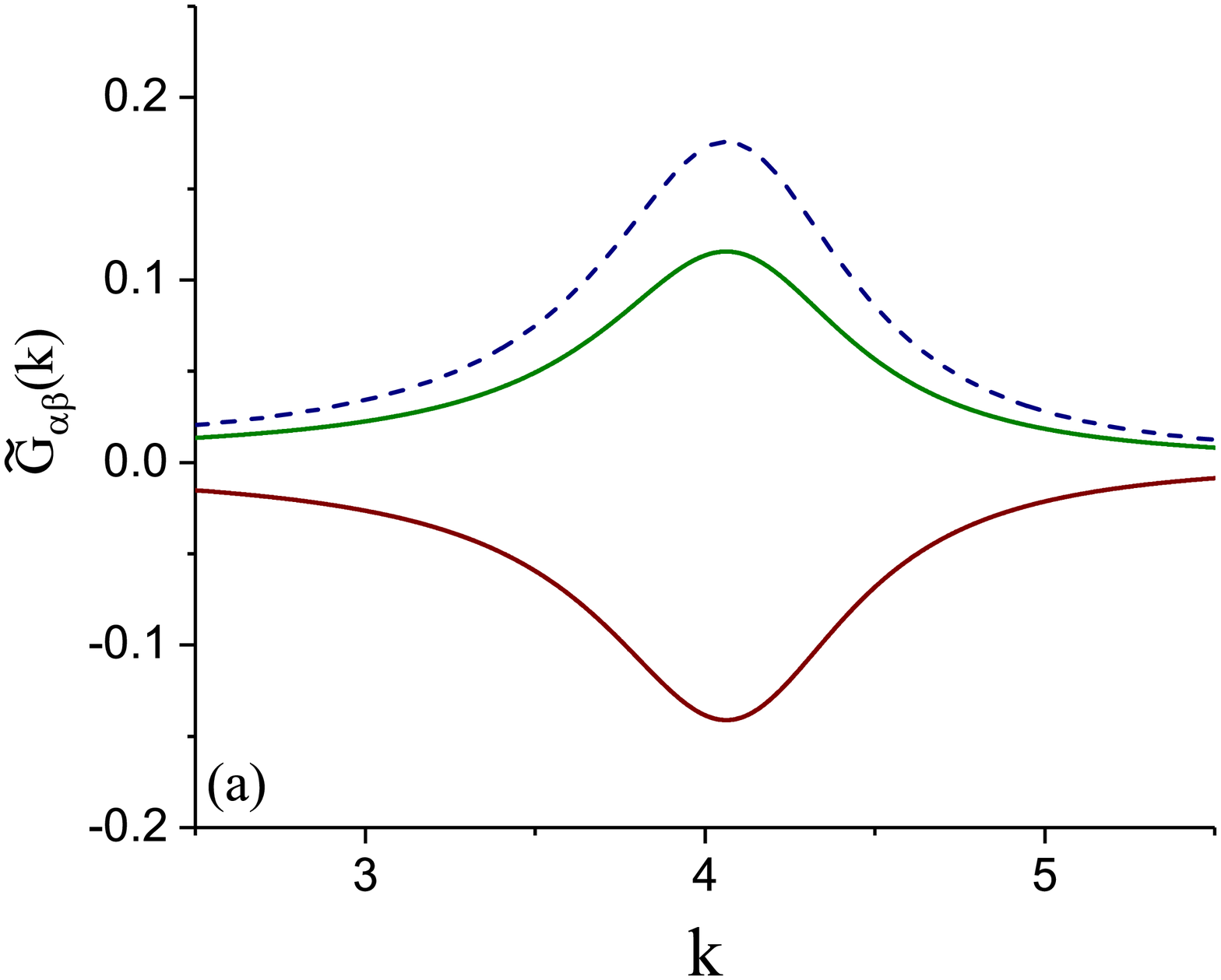} 
	\qquad
	\includegraphics[clip,width=0.42\textwidth,angle=0]{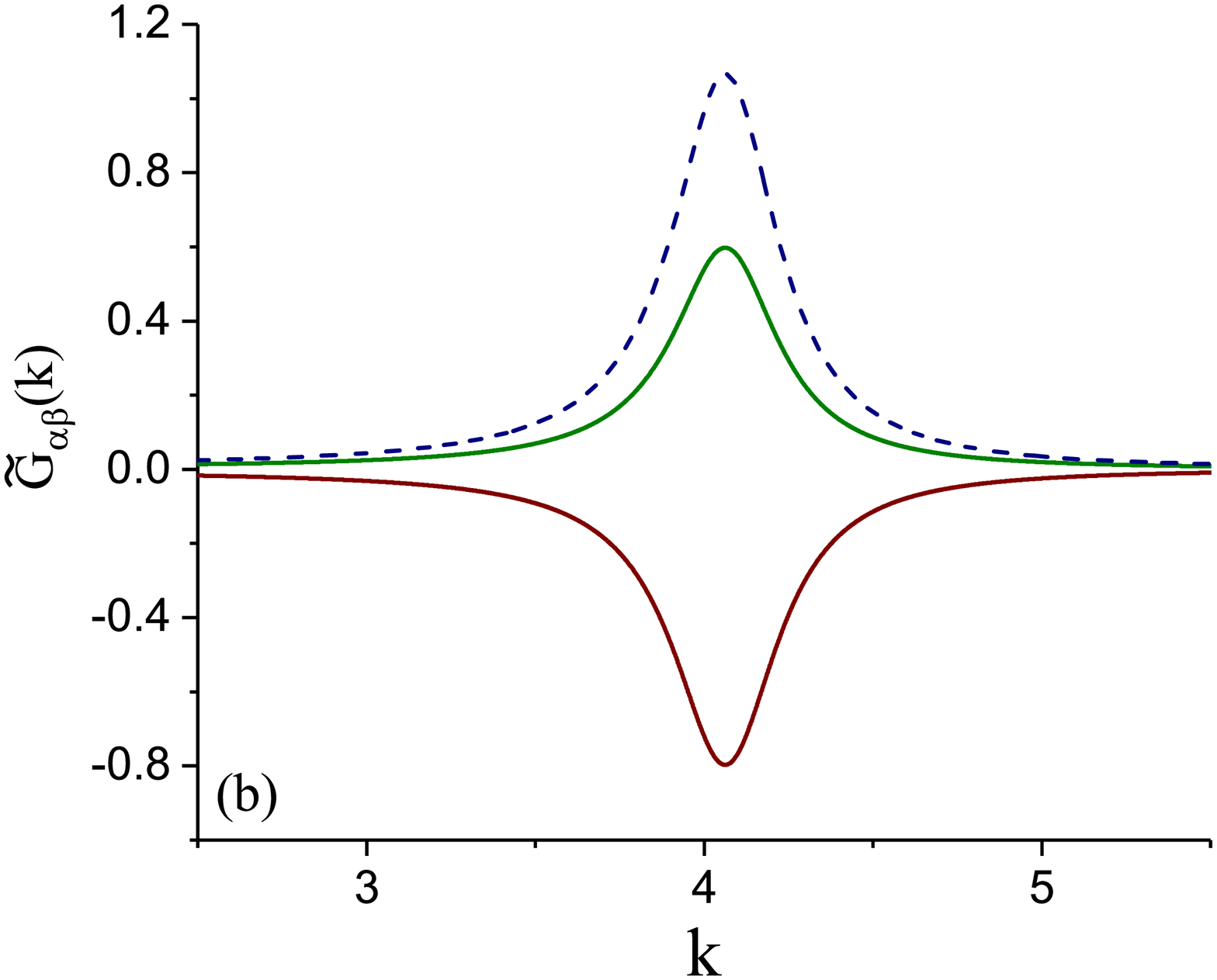} 
	\caption{\label{G_k_as06-fl} (Colour online) 
		Correlation functions   $\tilde G_{\alpha,\beta}(k)$  for  $\alpha=0.6$ 
		with the effect of fluctuations taken into account. Upper solid lines: $\tilde G_{11}(k)$, 
		dashed lines: $\tilde G_{22}(k)$ and lower solid lines: $\tilde G_{12}(k)$.  
		$T^*=0.1$, $\zeta=0.5$ and  $c=0.34 $ (panel~a), $c=0.54$  (panel~b).  
	}
\end{figure}

We focus on the region below the $\lambda$-surface  and calculate the correlation functions in Fourier representation taking 
into account fluctuations (Eqs.~(\ref{G22-fl-sw})-(\ref{D}) complemented by Eqs.~(\ref{C11(k0)})-(\ref{C12(k0)}) 
for $\tilde{C}_{\alpha\beta}(k_{0})$). We consider  the thermodynamic state denoted  by the filled square in  Fig.~\ref{spinodals_as06}.  The results   for $T^{*}=0.1$,  $\zeta=0.5$ and for two values of concentration, 
$c=0.34$ and $c=0.54$,  are presented in Fig.~\ref{G_k_as06-fl}. For 
$c>0.1$, a  maximum of $\tilde{G}_{22}(k)$ is higher than a maximum of  $\tilde{G}_{11}(k)$. 
The opposite situation is observed for $c<0.1$. 

\begin{figure}[h]
	\includegraphics[clip,width=0.6\textwidth,angle=0]{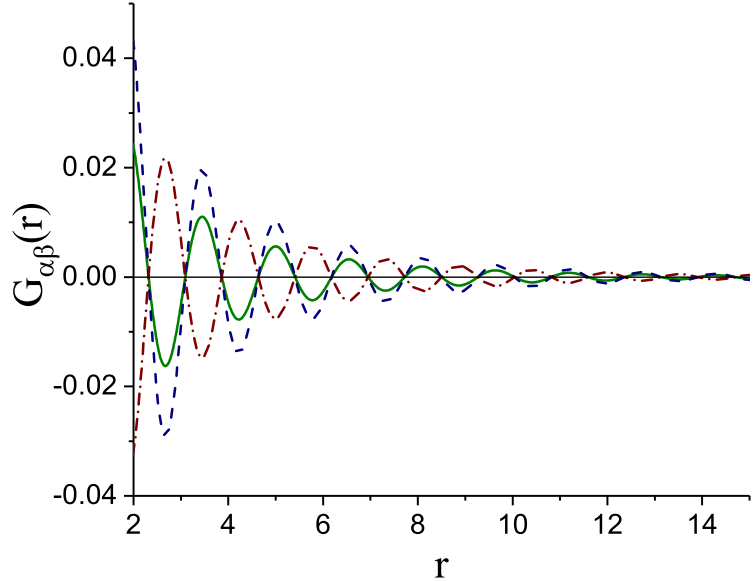}
	\caption{\label{GIJ_r_as06} (Colour online)
		Case $\alpha=0.6$. Correlation functions in real space  for $T^*=0.1$ ,  $\zeta=0.5$ and   $c=0.54$ with 
		the effect of fluctuations taken into account:
		$G_{11}(r)$ (solid line), $G_{22}(r)$ (dashed line) and $G_{12}(r)$ (dash-dotted line). $r$ is in $\sigma_{12}$ units.  
	}
\end{figure}

The correlation functions in real space for $T^*=0.1$,  $\zeta=0.5$ and   $c=0.54$ are presented in Fig.~\ref{GIJ_r_as06}. 
As for $\alpha=0.8$,  $G_{\alpha\beta}(r)$ have the form (\ref{Gr})
and  show exponentially damped oscillatory decay with the period of damped oscillations $\lambda/\sigma_{12}\simeq 1.54$. 
A maximum of  $G_{\alpha\alpha}(r)$ coincides with a minimum of $G_{12}(r)$.
  In the present case, however, the  oscillations are  more long-ranged. 
  For the considered thermodynamic state,  $\alpha_{0}\sigma_{12}\simeq 0.19$, $\alpha_{1}\sigma_{12}\simeq 4.1$, $A_{11}\simeq 0.075$,
  $A_{22}\simeq 0.13$, and $A_{12}\simeq -0.096$. As for $\alpha=0.8$, the amplitudes satisfy the  rule (\ref{AIJ}). 

\begin{figure}[h]
	\centering
	\includegraphics[clip,width=0.42\textwidth,angle=0]{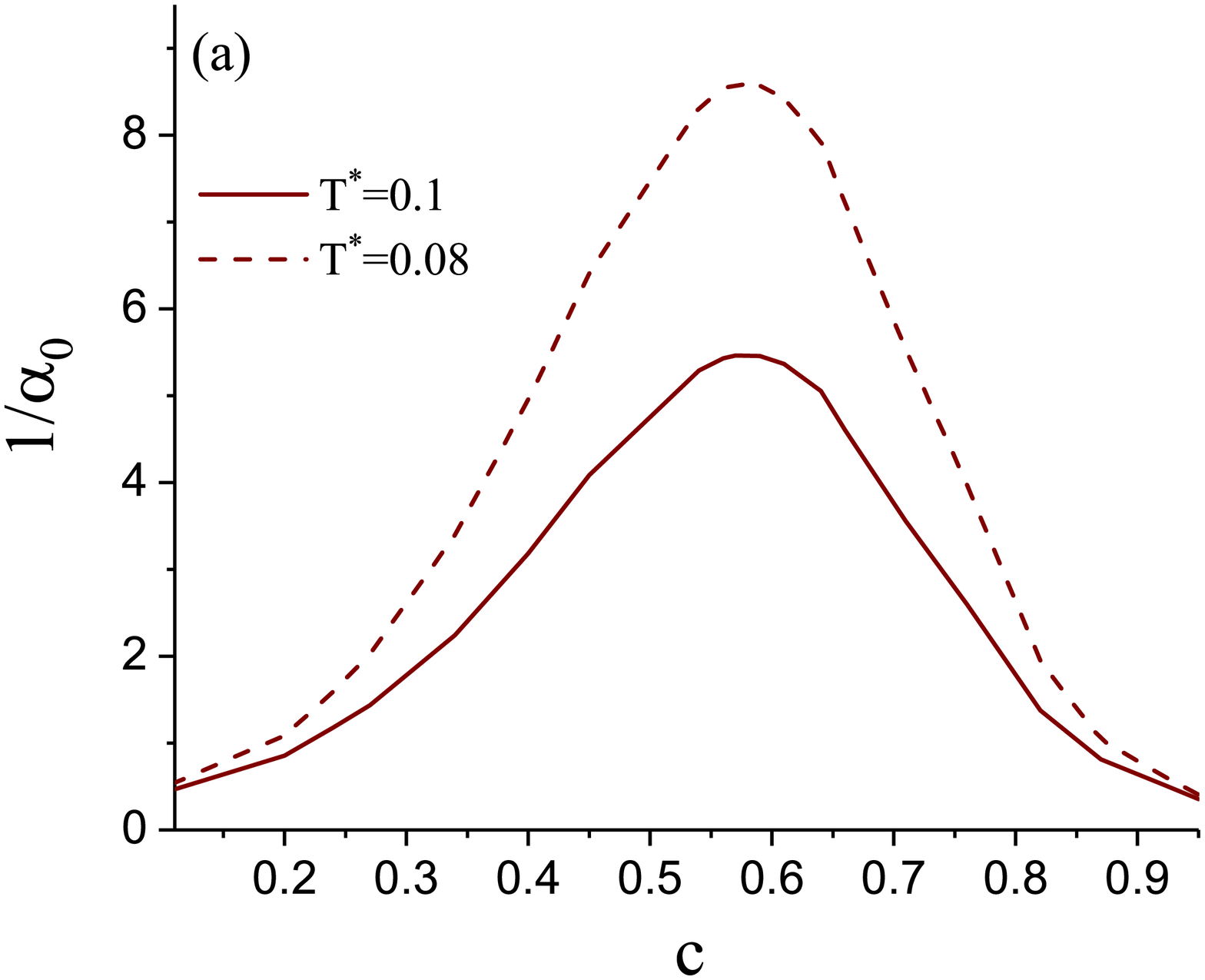}
	\qquad
	\includegraphics[clip,width=0.42\textwidth,angle=0]{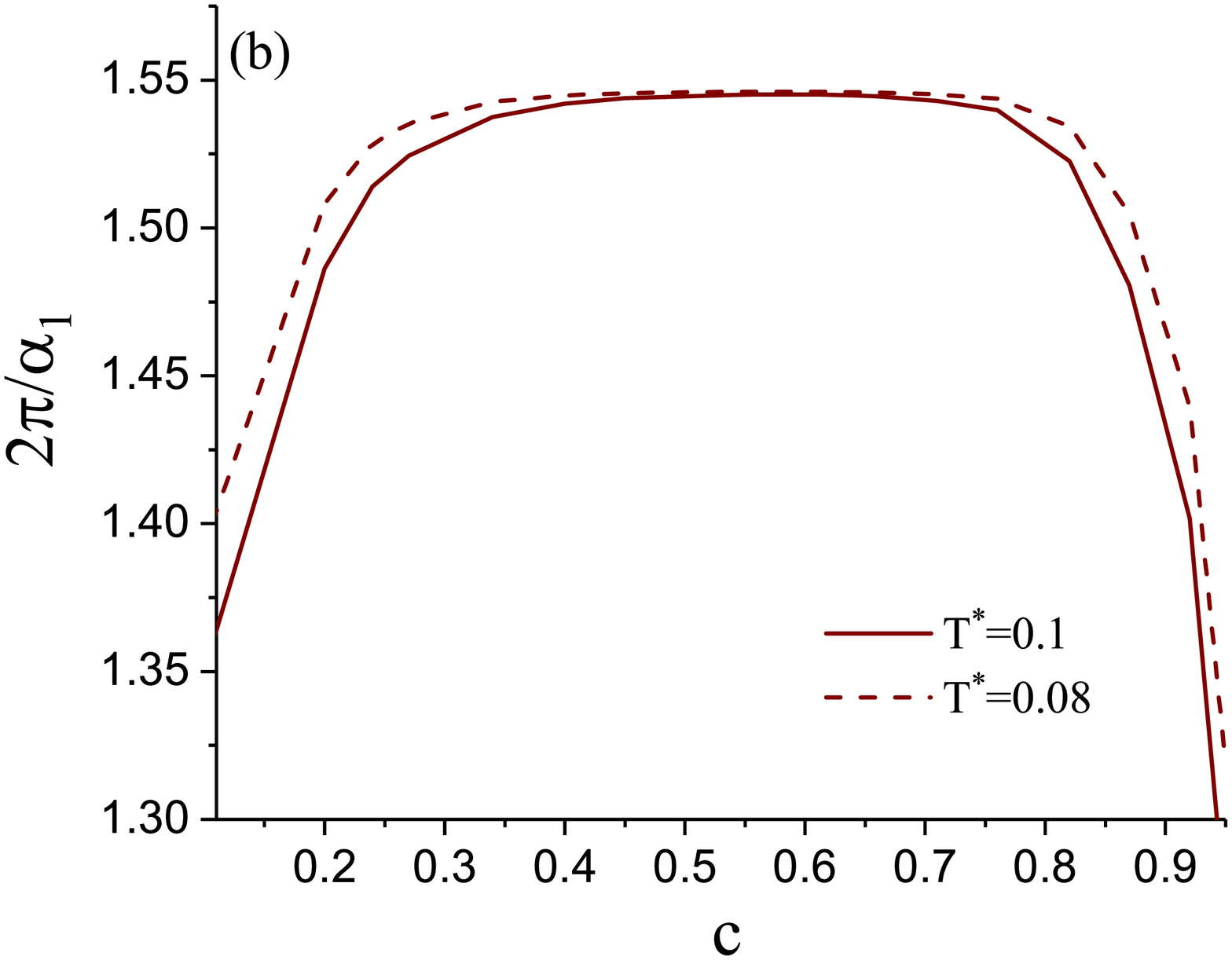}
	\caption{\label{xi-l-as06}
		The correlation length $1/\alpha_{0}$ (panel~a) and the period of oscillations $\lambda=2\pi/\alpha_{1}$
		(panel~b) of the correlation functions $G_{\alpha\beta}(r)$  for $\alpha=0.6$ as functions of the concentration $c$ for two values of temperature:  $T^{*}=0.08$ (dashed line)  and $T^{*}=0.1$ (solid line). $\alpha_{0}$ and $\alpha_{1}$ are in units of $\sigma_{12}^{-1}$. 
	}
\end{figure}

In Fig.~\ref{xi-l-as06},  the decay length  $\alpha_{0}^{-1}$ and the period of  damped oscillations $\lambda$  are presented as
functions of the concentration $c$ for
fixed total volume fraction ($\zeta=0.5$) and for two values of the temperature: $T^{*}=0.08$   and $T^{*}=0.1$.
In general, the dependence   of both quantities on $c$ is similar  to the behaviour observed for   $\alpha=0.8$. However,  for $\alpha=0.6$ the  decay length is larger than in the case $\alpha=0.8$ (for comparison, see also Table~\ref{Table2} where the results  are shown for $\zeta=0.4$ and $0.5$).

\subsubsection{Simulation results}
 The details of simulations for a binary mixture with $\alpha=0.6$ are given in Table~\ref{TableI} (see System~II ). 

\begin{figure}[h]
	\centering
	\includegraphics[clip,width=0.45\textwidth,angle=0]{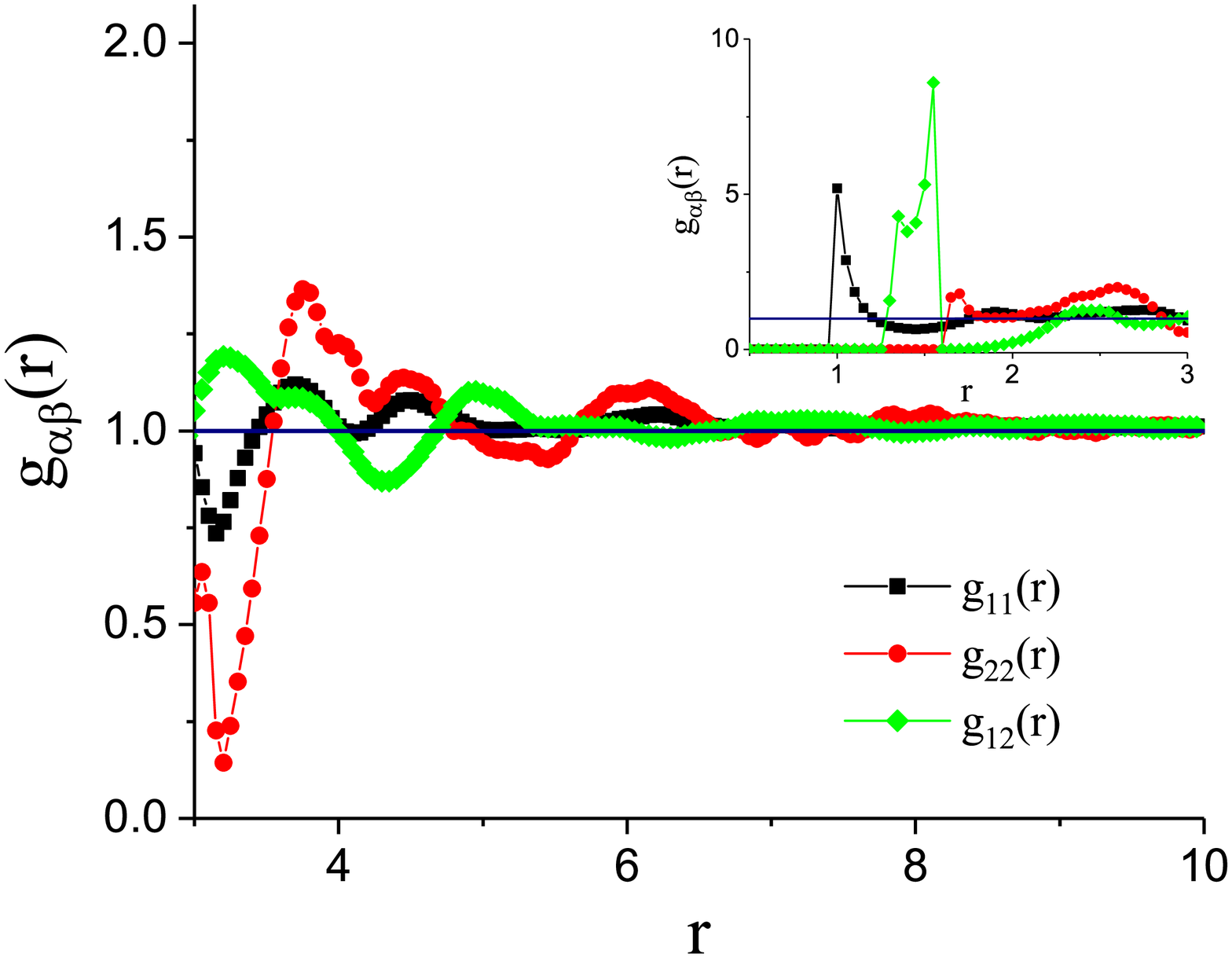}
	\quad
	\includegraphics[clip,width=0.38\textwidth,angle=0]{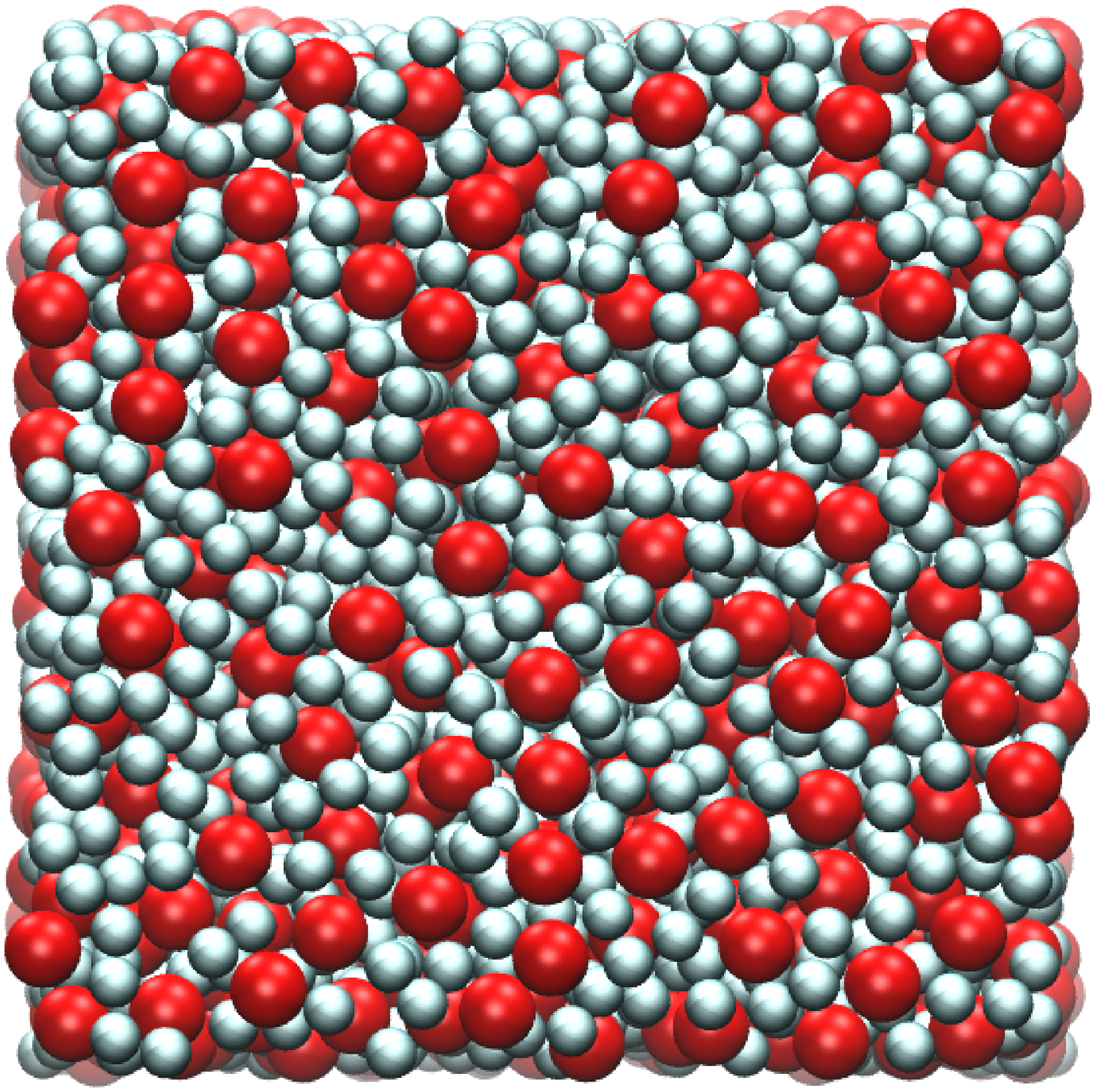}
	\caption{(Colour online)   Pair distribution functions and a representative configuration of the model with $\alpha=0.6$ for $T^*=0.1$, total volume fractions 
		$\zeta=0.5$ and concentration $c=0.54$.  $r$ is in $\sigma_{1}$ units.}
	\label{sim_sqw_as06-1}
\end{figure}

The pair distribution functions $g_{\alpha\beta}(r)$ and a representative configuration are presented in Fig.~\ref{sim_sqw_as06-1} 
for $T^*=0.1$, $\zeta=0.5$ and $c=0.54$. 
One can see an oscillatory decay of $g_{\alpha\beta}(r)$ with the period of damped oscillations $\lambda\approx 2\sigma_{1}$  that agrees with the theoretical results  (see Fig.~\ref{GIJ_r_as06}  noting that $\sigma_{12}\simeq 1.3\sigma_{1}$). The amplitude of $g_{22}(r)$ is  noticeably  larger than the amplitude of $g_{11}(r)$, that agrees with theoretical results for $G_{\alpha\beta}(r)$,  the minimum of $g_{\alpha\alpha}(r)$ coincides with the maximum of $g_{12}(r)$, and vice versa. 

One can see on the snapshot that the bigger particles are surrounded by the smaller particles, and stay apart from one another, except from very few cases.  
The distribution of the small particles within the attractive shell around the bigger particle is shown in Fig.~\ref{hist_as}
(panel~b). The most probable number of smaller particles in the attractive shell of the big one is $N_1=15$, 
but as in the case of $\alpha=0.8$,
the fluctuations of $N_1$ are large.

\begin{figure}[h]
	\centering	
	\includegraphics[clip,width=0.45\textwidth,angle=0]{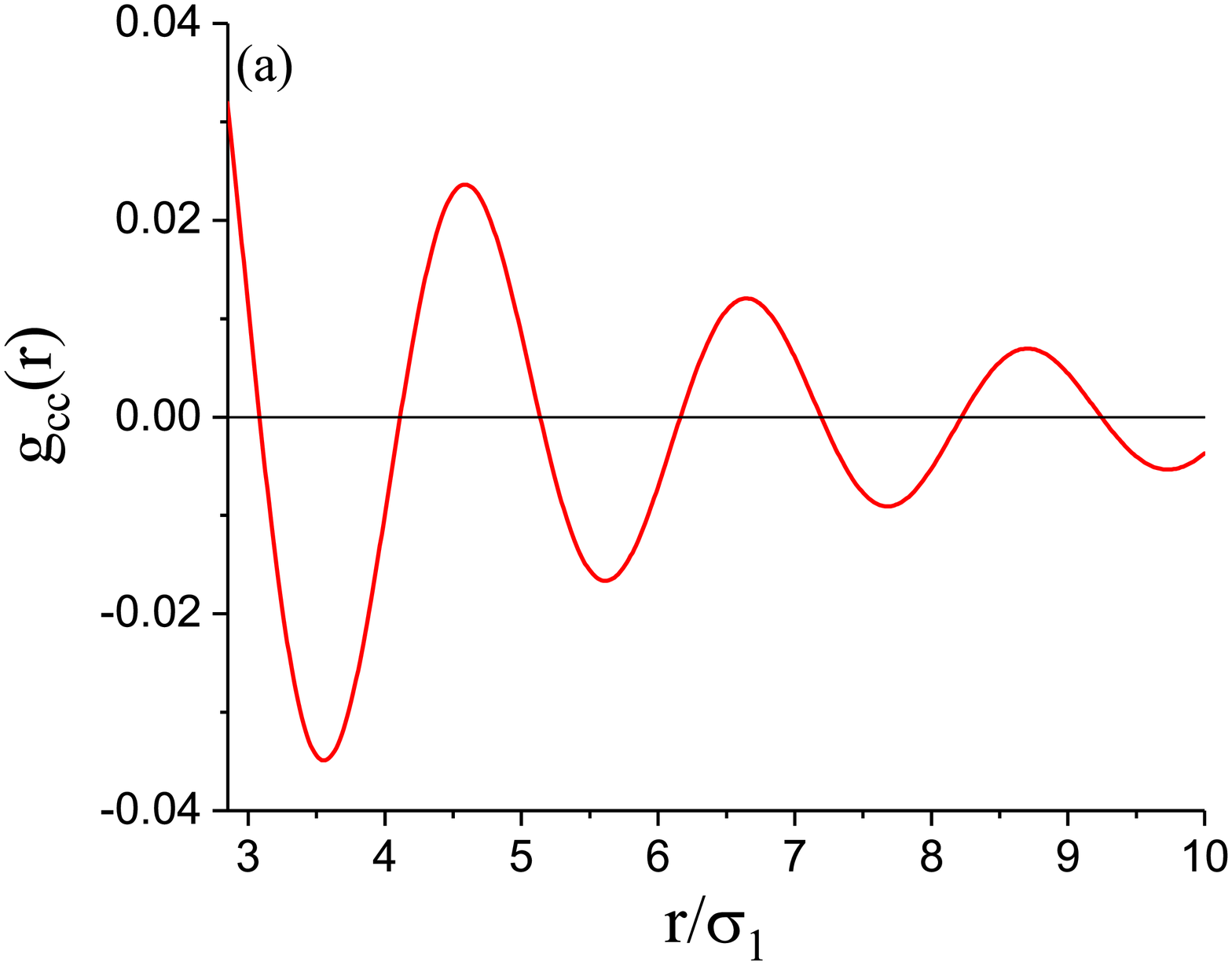} 
	\quad
	\includegraphics[clip,width=0.45\textwidth,angle=0]{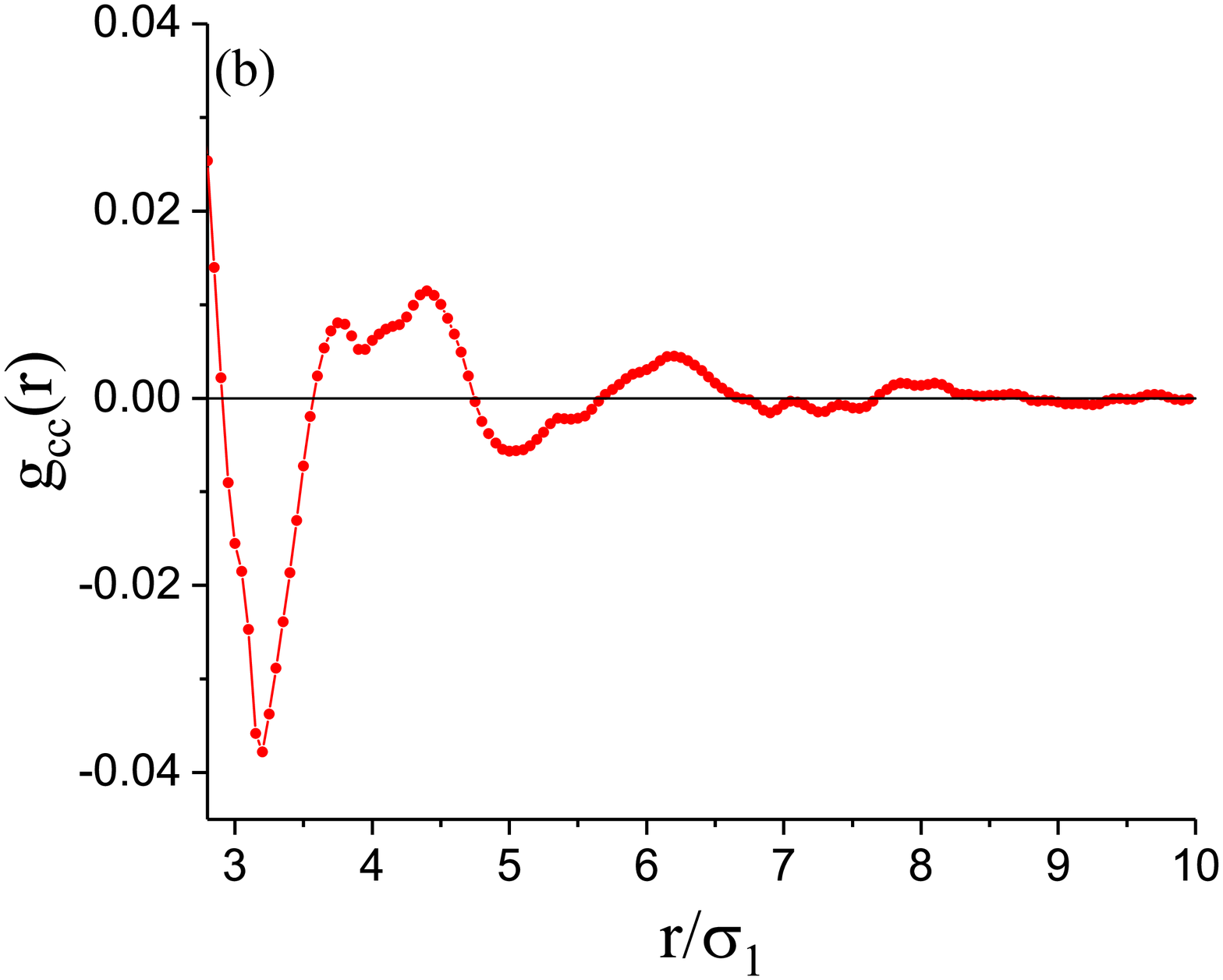}
	\caption{\label{g_cc_r_as06} (Colour online)  Concentration-concentration distribution function $g_{cc}(r)$ for $\alpha=0.6$ and for  $T^*=0.1$, 	$\zeta=0.5$ and
	  $c=0.54$.   Panel~a: theoretical results,  panel~b: simulation results.  $r$ is in $\sigma_{1}$ units.
	}
\end{figure}

The concentration-concentration distribution function
obtained from our theory and by MC simulations 
for $T^{*}=0.1$,  $\zeta=0.5$ and $c=0.54$ is shown in~Fig~\ref{g_cc_r_as06}. We can see that the theoretical and simulation 
results are in  reasonable agreement: in both cases the distribution function  $g_{cc}(r)$ shows oscillatory decay with the 
same period of damped oscillations: $\lambda\approx 2\sigma_{1}=1.5\sigma_{12}$. The amplitudes are in semiquantitative agreement, 
but the correlation length obtained in the theory is larger.

\subsection{Case $\alpha=0.25$} 
\subsubsection{Theoretical results} 
\begin{figure}[h]
	\centering
	\includegraphics[clip,width=0.5\textwidth,angle=0]{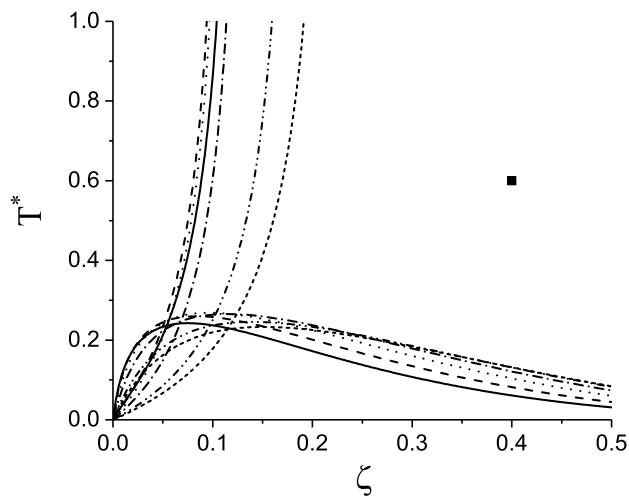}
		\vspace{-5mm}		
	\caption{\label{spinodals_as025}
		Gas-liquid spinodals and $\lambda$-lines for the model (\ref{V_sq-well}) at $a=1.2$ and $\alpha=0.25$ 
		for different values of concentration. Solid lines: $c=0.62$  ($x=0.025$), dashed lines: $c=0.77$  ($x=0.05$), 
		dotted lines: $c=0.88$ ($x=0.1$), dash-dotted lines: $c=0.94$ ($x=0.2$), dash-dot-dotted lines: $c=0.98$ ($x=0.4$), 
		and short-dashed lines: $c=0.99$ ($x=0.5$). The filled square denotes the thermodynamic state located below the 
		$\lambda$-surface but above the gas-liquid spinodal (see  text for more details).
	}
\end{figure}

Finally, we focus on a large size asymmetry of the particles of different species, namely, we consider a binary mixture with  
$\alpha=0.25$ ($\sigma_{2}=4\sigma_{1}$). In this case, the MF boundaries of stability  
for the fixed values of concentration have the form presented  in Fig.~\ref{spinodals_as025}. As for a smaller size asymmetry,  
the dependence of $T^*$ on  the concentration  at the fixed  $\zeta$ is nonmonotonic. 
In addition, the form of the $\lambda$-lines changes significantly:  starting from a certain value of $\zeta$, they  become 
almost parallel to the $y$-axis.
\begin{figure}[h]
	\centering
	\includegraphics[clip,width=0.42\textwidth,angle=0]{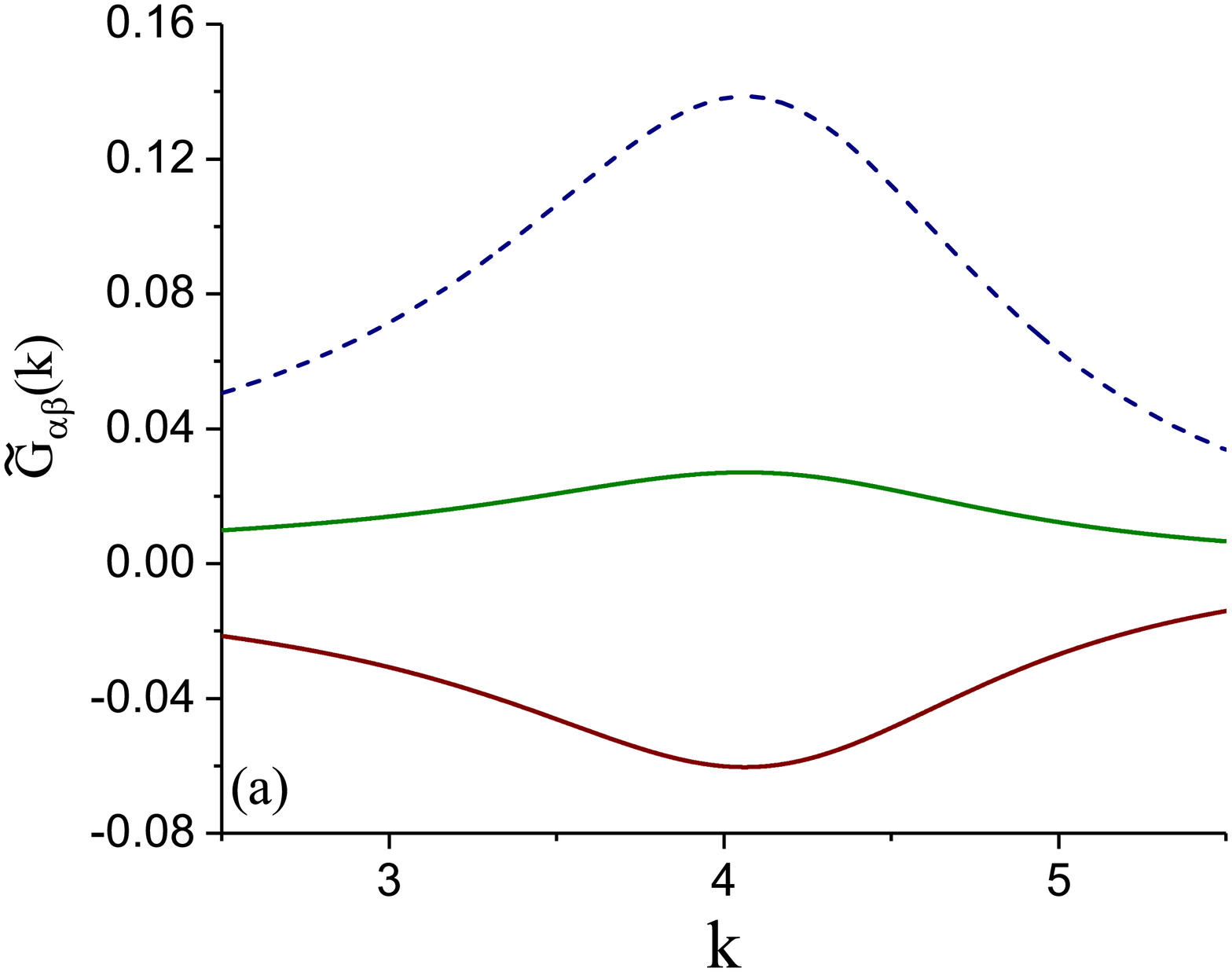} \qquad
	\includegraphics[clip,width=0.42\textwidth,angle=0]{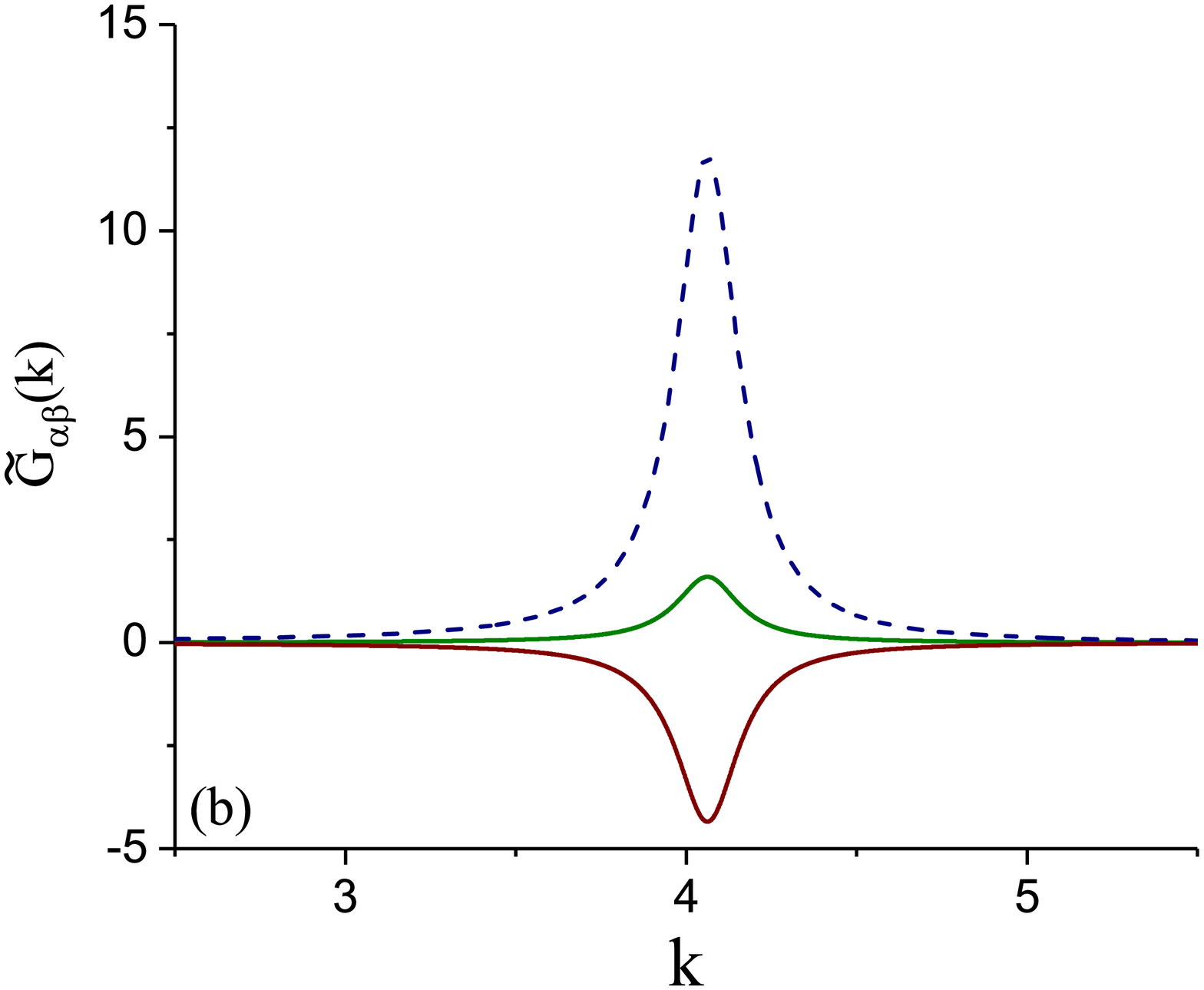}
	\caption{\label{G_k_as025-fl} (Colour online)
		Correlation functions   $\tilde G_{\alpha,\beta}(k)$  for $\alpha=0.25$ 
		with the effect of fluctuations taken into account. Upper solid lines: $\tilde G_{11}(k)$, 
		dashed lines: $\tilde G_{22}(k)$ and lower solid lines: $\tilde G_{12}(k)$.  $T^*=0.6$, $\zeta=0.4$ and  $c=0.49$ 
		(panel~a),  $c=0.67$  (panel~b).
	}
\end{figure}

We calculate the correlation functions in Fourier representation  for $T^*=0.6$, $\zeta=0.4$ and for two values of  the concentration of 
the big particles: $c=0.49$  and $c=0.67$ (the corresponding thermodynamic state is denoted by the filled square in 
Fig.~\ref{spinodals_as025}).  
The results are presented in Fig.~\ref{G_k_as025-fl}. For  $c=0.67$, the dependences of 
$\tilde G_{\alpha\beta}(k)$  on $k$ show narrow peaks at $k=k_{0}$ (Fig.~\ref{G_k_as025-fl}, panel~b) which widen  
already  for $c=0.62$ and simultaneously the heights of the peaks reduce. 
For $c=0.49$, the extrema  of $\tilde G_{\alpha\beta}(k)$ become flat (Fig.~\ref{G_k_as025-fl}, panel~a).  
It should be noted  that for the considered thermodynamic states,  
Eqs.~(\ref{k0})-(\ref{C12(k0)}) have no solutions for $c\ge 0.7$.

\begin{figure}[h]
	\includegraphics[clip,width=0.5\textwidth,angle=0]{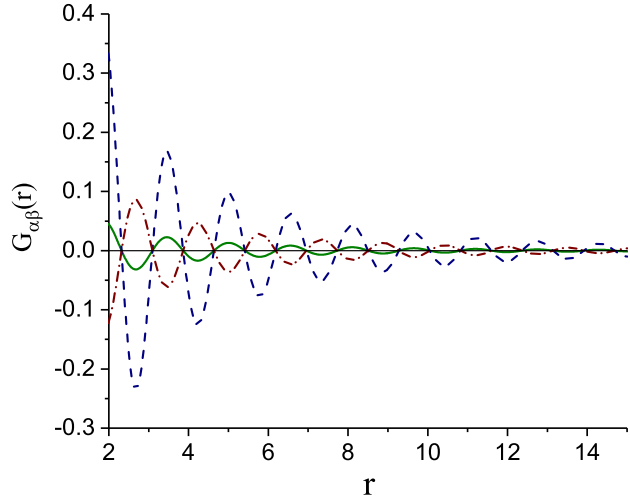}
	\caption{\label{GIJ_r_as025} (Colour online)
		Case $\alpha=0.25$. Correlation functions in real space  with the effect of fluctuations taken into account for  $T^*=0.6$,  $\zeta=0.4$ and   $c=0.67$: 
		$\tilde G_{11}(k_0)$ ( solid line), $\tilde G_{22}(k_0)$ (dashed line) and $\tilde G_{12}(k_0)$ (dash-dotted line). 
		$r$ is in $\sigma_{12}$ units.
	}
\end{figure}

 In Fig.~\ref{GIJ_r_as025}, the correlation functions in real space representation are  presented
 for  $T^*=0.6$,  $\zeta=0.4$ and  $c=0.67$. As for a smaller size asymmetry, the correlation functions $G_{\alpha\beta}(r)$ 
 have the form (\ref{Gr}) and  show exponentially damped oscillatory decay with the period of damped oscillations 
 $\lambda\approx 1.55\sigma_{12}$. For $T^*=0.6$,  $\zeta=0.4$ and  $c=0.67$, we have
  $\alpha_{0}\sigma_{12}\simeq 0.11$, $\alpha_{1}\sigma_{12}\simeq 4.1$. The amplitudes obey the rule  Eq.~(\ref{AIJ}) 
  and their values are as follows: $A_{11}\simeq 0.12$, $A_{22}\simeq 0.85$, and $A_{12}\simeq -0.32$. 

\begin{figure}[h]
	\centering
	\includegraphics[clip,width=0.4\textwidth,angle=0]{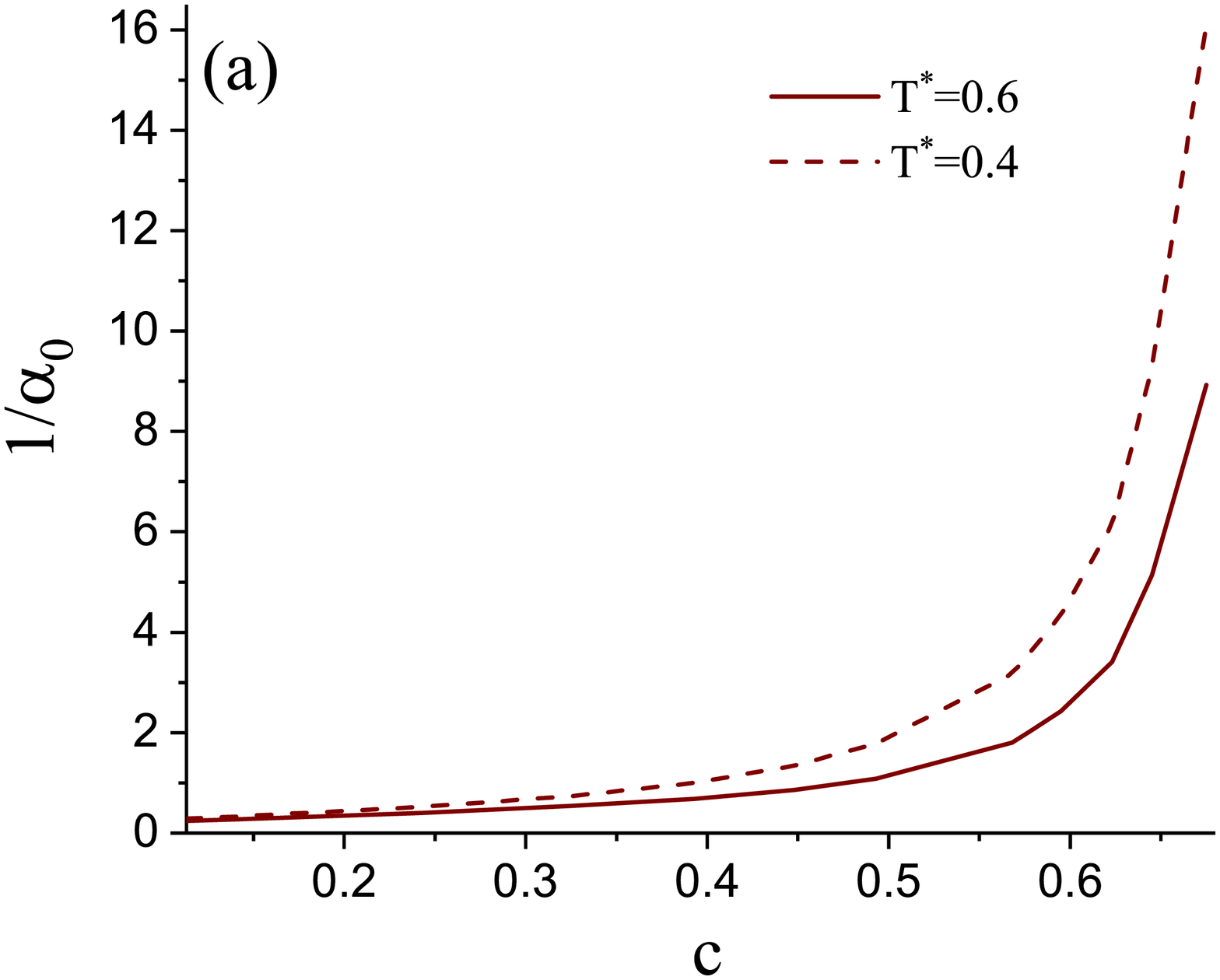}
	\qquad
	\includegraphics[clip,width=0.4\textwidth,angle=0]{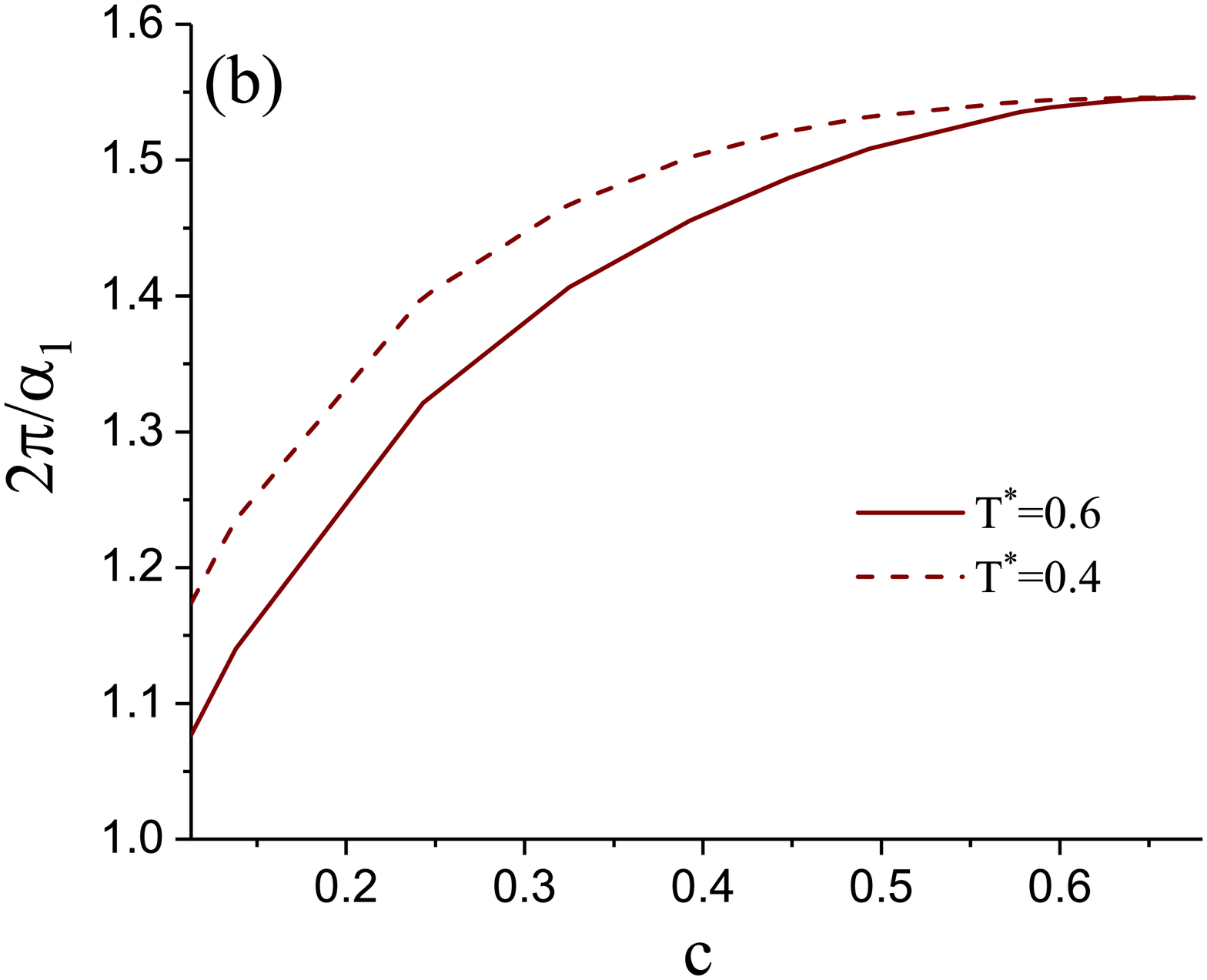}
	\caption{\label{xi-l25}
		The decay length $\alpha_{0}^{-1}$ (panel~a) and the period of oscillations $\lambda=2\pi/\alpha_{1}$ 
		(panel~b) of the correlation functions $G_{\alpha\beta}(r)$ for $\alpha=0.25$ as 
		functions of the concentration  for  two values of temperature: $T^{*}=0.4$ (dashed curve)  and $T^{*}=0.06$ (solid curve). 
		$\alpha_{0}$ and $\alpha_{1}$ are in units of $\sigma_{12}^{-1}$. 
	}
\end{figure}

In Fig.~\ref{xi-l25},   the decay length  $\alpha_{0}^{-1}$ and the period of  damped oscillations $\lambda$ are presented  
as functions of the concentration $c$ for the fixed total volume fraction ($\zeta=0.4$) and for two values of the temperature: $T^{*}=0.4$   and $T^{*}=0.6$. In contrast to the cases of a  smaller size asymmetry,    the decay length first increases very slowly on increasing the concentration  and then   (for $c>0.55$) reaches its maximum very rapidly.
The largest value of $\alpha_{0}^{-1}$  increases significantly with  decreasing temperature. 
Moreover, the decay length for $\alpha=0.25$ is much larger than in the case of  smaller size asymmetry, even for the  higher temperature (see the data in Table~\ref{TableI}).
 $\lambda$, in turn, does not exhibit a steep increase  with $c$. 

\subsubsection{Simulation results} 
The simulations are performed for   $T^*=0.6$, $\zeta=0.4$ and   $c=0.67$.
The  details of simulation  are presented in Table~\ref{TableI} (System III). 
In this case, however, we limit ourselves to determination of the propertis of the attractive shell of the big particles. 
The previous cases confirm that the distribution of the particles on the mesoscopic length scale can be predicted by our theory
with semi-quantitative agreement with simulations. Thus, it is not necessary to determine the long-distance properties of 
the pair-distribution function by simulations that for large size asymmetry are much more demanding~\cite{ashton:10:0}.
On the other hand, the short-distance structure cannot be determined by our mesoscopic theory, and this complementary information about 
local ordering can be obtained by our simulation procedure. 
\begin{figure}[h]
	\centering
	\includegraphics[clip,width=0.55\textwidth,angle=0]{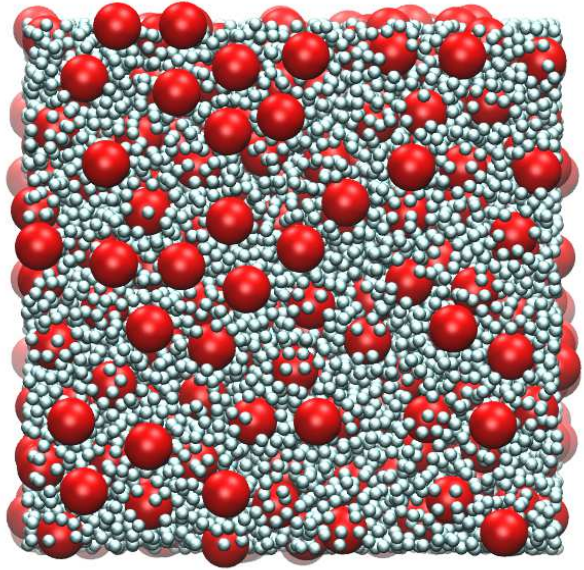}
	\caption{(Colour online) A representative configuration  of the model with $\alpha=0.25$ for $T^*=0.6$,  total volume fraction
		$\zeta=0.4$ and concentration $c=0.67$.}
	\label{sim_sqw_as025-2}
\end{figure}

In Fig.~\ref{sim_sqw_as025-2}, a representative configuration is shown. 
As it is seen from the snapshot, the big particles are surrounded by the small particles, and have a strong tendency 
to the periodic ordering. This is also supported by the histogram in  Fig.~\ref{hist_as} (panel~c).
As one can see,  $P(N_{1})$ is quite narrow in contrast to the previously considered cases. The ratio between
the standard deviation 
and the average number of the 'adsorbed' particles is quite small, and can be estimated 
as $\sim 5\%$. The theoretical results show quite large correlation length for the considered thermodynamic state.
We conclude that the periodic ordering on the length scale of $~\sigma_2$, and the ordering near the surface of the big particles
(meaning large number of the small particles in the attractive shell and small fluctuations of this number) go together.

\section{Summary and Conclusions}

We have studied binary mixtures with effective interactions between the particles that favour nearest-neighbours of different kind.
At low temperature, such interactions lead to a periodic structure with alternating  particles of the first 
and the second component. At higher $T$ the crystal melts, but a 
competition between the ordering effect of energy and disordering effect of entropy 
leads to local periodic order. Our aim was determination of the effect of size asymmetry of the particles of 
the two species on this local order.

We have calculated correlation functions for small, medium and large size asymmetry within the mesoscopic DFT developed in
Ref.\cite{ciach:11:2,Ciach:20:1}, at the lowest nontrivial order 
beyond MF. 
Our results were favourably compared with MC simulations on a semiquantitative level. In addition, 
the simulations allowed us to determine the distribution of the small particles in the attractive shell of the bigger ones.

We have found that for all considered size ratios $\alpha$ the correlation functions show exponentially damped oscillations, 
with the maxima of the correlation
function for like species 
that coincide with the minima of the correlation function for different species.
The range and amplitude of the correlations increase with decreasing $T$ and/or increasing total
volume fraction of the particles.  In each considered system, the period $2\pi/\alpha_1$ of the damped oscillations is almost independent 
of their relative volume fraction of the large particles, $c=\zeta_2/\zeta$, as long as the correlation length is larger than $1$ 
(in $\sigma_{12}$ units).
Moreover, $\alpha_1$  is very close to the wave-number corresponding 
to the first maximum of the interaction between
different species in Fourier representation. 
Only for the correlation length $1/\alpha_0<1$, i.e. too small for formation of the periodic structure on the mesoscopic length scale, 
the period of the damped oscillations decreases.
Another common feature of all the considered cases is the fact that  the correlations between the bigger particles are the strongest 
for a large interval of their relative volume fraction $c$.
Only for $c$ smaller than a value depending on $\alpha$, the correlations between the smaller particles are larger.
This value of  $c$, however, decreases rapidly  with decreasing $\alpha$ (increasing size ratio).

In addition to the above  similarities, 
there are significant differences between the properties of the systems with different size ratio $\alpha$. 
First of all, correlations increase significantly with increasing size ratio, and 
are much stronger in the case of $\alpha=0.25$ than for $\alpha\ge 0.6$. The amplitude and range of correlations in the case
of large size-asymmetry  are significantly larger 
even at $T$ much higher than in the system with moderate size ratio. 
The dependence of the local order on $c$ in the system with large  size-ratio is qualitatively different than in the systems with
moderate or small size asymmetry. 
In the latter two cases, the correlation length takes a pronounced maximum for comparable volume fractions of the two species.
With increasing size asymmetry,
the maximum of the correlation length moves to a larger volume fraction of the bigger particles. 
The difference between the amplitudes of the correlation function between the big particles and the remaining correlation
functions increases with increasing size asymmetry too.
For $\alpha=0.25$, however, the correlation length is very small for small $c$, 
and increases very slowly with $c$ for $c<0.5$. For $c>0.5$ a rapid increase to large numbers occurs, and both $1/\alpha_0$ 
and $2\pi/\alpha_1$ are monotonic functions of $c$, in contrast to the previous two cases. Moreover, the magnitude of the correlations 
between the big particles becomes much larger than the magnitude of the remaining correlations for all values of $c$ 
corresponding to $1/\alpha_0>1$.

The stronger ordering on the mesoscopic length
scale is accompanied
by stronger ordering in the attractive shell of the big particles. For $\alpha=0.8$ and $\alpha=0.6$,
the average number of the smaller particles 
located in the attractive shell of the big particle is $8$ and $15$, respectively.  This number, however, fluctuates strongly, and
in the considered thermodynamic state, the ratio between the standard deviation and the average number of particles can 
be estimated as $\sim 25\%$ and $\sim 20\%$, respectively. 
On the other hand, for $\alpha=0.25$ and the considered thermodynamic state, the average number of the
small particles inside the attractive shell of the big one 
is $37.5$, and the ratio between the standard deviation and the average number of particles can be estimated as $\sim 5\%$. 
Notably, the temperature in this case is much higher than in the above two cases.

Our results show that for all considered cases, the bigger particles are distributed much more uniformly than in the random distribution.
Even though the long-range order is lacking, the distance between the nearest-neighbours of the same kind is approximately the same,
and the big particles are separated by the small ones
(see Fig.\ref{sim_sqw_as025-2}).
In some way our models resemble hyperuniform systems~\cite{torquato:03:0,meyra2019hyperuniformity}. 
Our results suggest that the larger is the size ratio, the more uniformly are distributed the larger
particles.

\end{document}